\documentclass[twocolumn, tighten]{aastex631}

\usepackage{float}
\usepackage{amssymb, amsmath}

\newcommand{\pgc}{PGC~66551}
\newcommand{\Ms}{$\textrm{M}_{\odot}$}
\newcommand{\Msperyr}{\textrm{M}$_{\odot}$~\textrm{yr}$^{-1}$}
\newcommand{\kms}{$\textrm{km~s$^{-1}$}$}
\newcommand{\oi}{\,{\sc i}}
\newcommand{\ii}{\,{\sc ii}}
\newcommand{\iii}{\,{\sc iii}}
\newcommand{\Ha}{H$\alpha$}
\newcommand{\Hb}{H$\beta$}
\newcommand{\Hg}{H$\gamma$}

\newcommand{\nb}{\textsc{NBursts}}
\newcommand{\mgb}{Mg\textit{b}}
\newcommand{\av}{$A_V$}

\graphicspath{{./}{figs/}}

\received{April 18, 2023}
\revised{November 20, 2023}
\accepted{November 21, 2023}

\submitjournal{ApJ}

\begin{document}

\title{Probing the History of the Galaxy Assembly of the Counter-rotating Disk Galaxy PGC~66551}

\correspondingauthor{Ivan Katkov}
\email{ivan.katkov@nyu.edu}

\author[0000-0002-6425-6879]{Ivan Yu. Katkov}
\affiliation{New York University Abu Dhabi, P.O. Box 129188, Abu Dhabi, UAE}
\affiliation{Center for Astrophysics and Space Science (CASS), New York University Abu Dhabi, PO Box 129188, Abu Dhabi, UAE}
\affiliation{Sternberg Astronomical Institute, Lomonosov Moscow State University, Universitetskij pr., 13,  Moscow, 119234, Russia}

\author[0000-0002-1750-2096]{Damir Gasymov}
\affiliation{Faculty of Physics, Lomonosov Moscow State University, 1 Leninskie Gory, Moscow 119991, Russia}
\affiliation{Sternberg Astronomical Institute, Lomonosov Moscow State University, Universitetskij pr., 13,  Moscow, 119234, Russia}

\author[0000-0001-8646-0419]{Alexei Yu. Kniazev}
\affiliation{South African Astronomical Observatory, P.O. Box 9, 7935 Observatory, Cape Town, South Africa}
\affiliation{Southern African Large Telescope Foundation, P.O. Box 9, 7935 Observatory, Cape Town, South Africa}
\affiliation{Sternberg Astronomical Institute, Lomonosov Moscow State University, Universitetskij pr., 13,  Moscow, 119234, Russia}
\affiliation{Special Astrophysical Observatory, Nizhnij Arkhyz, Karachai-Circassia, 369167, Russia}

\author[0000-0003-4679-1058]{Joseph D. Gelfand}
\affiliation{New York University Abu Dhabi, P.O. Box 129188, Abu Dhabi, UAE}
\affiliation{Center for Astrophysics and Space Science (CASS), New York University Abu Dhabi, PO Box 129188, Abu Dhabi, UAE}
\affiliation{Center for Cosmology and Particle Physics, New York University, 726 Broadway, room 958, New York, NY 10003}

\author[0000-0001-8427-0240]{Evgenii V. Rubtsov}
\affiliation{Sternberg Astronomical Institute, Lomonosov Moscow State University, Universitetskij pr., 13,  Moscow, 119234, Russia}

\author[0000-0002-7924-3253]{Igor V. Chilingarian}
\affiliation{Center for Astrophysics---Harvard and Smithsonian, Cambridge, USA}
\affiliation{Sternberg Astronomical Institute, Lomonosov Moscow State University, Universitetskij pr., 13,  Moscow, 119234, Russia}

\author[0000-0003-4946-794X]{Olga K. Sil'chenko}
\affiliation{Sternberg Astronomical Institute, Lomonosov Moscow State University, Universitetskij pr., 13,  Moscow, 119234, Russia}



\begin{abstract}

Stellar counter-rotation in disk galaxies directly relates to the complex phenomenon of the disk mass assembly believed to be driven by external processes, such as accretion and mergers.
The detailed study of such systems makes it possible to reveal the source of external accretion and establish the details of this process.
In this paper, we investigate the galaxy PGC~66551 (MaNGA ID~1-179561) which hosts two large-scale counter-rotating (CR) stellar disks identified in the Sloan Digital Sky Survey MaNGA data and then confirmed using deep follow-up spectroscopy 
with the 10m Southern African Large Telescope.
We measured the properties of ionized gas and stellar populations of both CR disks in PGC~66551.
We found that the CR disk is compact, contains young stars with subsolar metallicity, and has a stellar mass of $5\times10^{9}$~\Ms\ which amounts to $\approx$20\% of the galaxy's total.
Surprisingly, the main 8~Gyr old disk has a significantly lower metallicity of $-0.8$~dex than other CR galaxies.
We developed a simple analytic model of the history of the metal enrichment, which we applied to PGC~66551 and constrained the parameters of the galactic outflow wind, and estimated the metallicity of the infalling gas that formed the CR disk to be $-0.9 ... -0.5$~dex.
Our interpretation prefers a merger with a gas-rich satellite over cold accretion from a cosmic filament as a source of gas, which then formed the CR disk in PGC~66551.

\end{abstract}



\section{Introduction} \label{sec:intro}

One of the fundamental questions in galaxy physics is ``how do galactic disks grow?''
External gas accretion is one of the important processes governing disk growth.
Many observational phenomena cannot be explained without gas accretion \citep[see][for a recent review]{Combes2014ASPC..480..211C, Putman2017ASSL..430....1P}.
For example, star formation rates (SFRs) observed at high redshifts suggest a galaxy would run out of gas in a few gigayears at most.
As a result, without external replenishment, most galaxies today would be \textit{red and dead}, contradicting the substantial star formation (SF) activity observed in most low-redshift systems \citep{Genzel2010MNRAS.407.2091G, Kennicutt1998ApJ...498..541K}.
It also turns out that the total budget of cold ($<10^4$~K) atomic gas present in galaxies does not decrease significantly as the stellar mass density evolves, which is expected to be due to the lack of accretion \citep{Madau2014ARA&A..52..415M, Neeleman2016ApJ...818..113N}.
Accretion of gas onto a galaxy is also required by chemical evolution models to successfully reproduce the metallicity distribution in the Milky Way and other galaxies \citep{Chiappini2009IAUS..254..191C, Gallazzi2005MNRAS.362...41G}.
Despite the importance of this process, there is a lot we do not know about it, for example, How much material can be acquired during accretion? How long does the accretion last and at what redshift does it typically occur? What are the sources of  acquired material? And finally, how does this process shape galaxies?
Furthermore, the direct detection as  well  as  the clear consequences of the past accretion are very elusive.
Hence, studies of galaxies with clear evidence of past accretion are very important.
Some of the best examples of such systems are galaxies with two counterrotating (CR) stellar disks.

CR stars are thought to originate from the infall of external material, the sources of which might be (i) cosmological filaments supplying pristine (i.e., primodial or very-low metallicity) gas \citep{Algorry2014MNRAS.437.3596A} or (ii) minor mergers with gas-rich satellites \citep{Thakar1996ApJ...461...55T, Thakar1998ApJ...506...93T,Lu2021MNRAS.503..726L}, followed by a burst of SF \citep{Pizzella2004A&A...424..447P}.
For the new gas with misaligned angular momentum to have settled in the main plane of the galaxy, the preexisting gas must have been removed, for example, by means of strong active galactic nuclei (AGN) feedback or gas stripping in a dense environment \citep{Starkenburg2019ApJ...878..143S}. 
Another possibility is that the infalling gas removes the preexisting interstellar medium (ISM) and then forms CR stars \textit{in situ} \citep{khoperskov_illustris}.
Both scenarios will lead to different time lags in SF and age differences of stellar population components.

All galaxies accrete material in some stage of their evolution, but in most cases, we cannot distinguish between accreted and original material in the finally shaped galaxy because the angular momenta of the accreted and original gas are sufficiently similar so that we cannot kinematically separate these two components.
However, this is not the case in CR galaxies, where they are sufficiently different to be able to be identified well after the merger occurred.
Kinematical disentanglement of the two disks provides a unique opportunity to probe stellar populations formed from the accreted material and the preexisted stellar populations before the infall event.

The presence of two CR stellar disks can be determined by a symmetric double peak on the velocity dispersion map, the so-called two-$\sigma$ feature \citep{Krajnovic2011MNRAS.414.2923K, Rubino2021A&A...654A..30R}.
Thanks to this simple diagnostics, numerous CR galaxies have recently been detected in the SDSS Mapping Nearby Galaxies at Apache Point Observatory (MaNGA) survey data \citep{Bao2022ApJ...926L..13B, Bevacqua2022MNRAS.511..139B}, exhibiting a variety of kinematic properties.
However, so far only about a dozen CR galaxies have been studied in detail so far using deep spectroscopic observations \citep{Coccato2011MNRAS.412L.113C, Coccato2013A&A...549A...3C, Coccato2015A&A...581A..65C, Katkov2013ApJ...769..105K, Katkov2016MNRAS.461.2068K, Johnston2013MNRAS.428.1296J}, and analyzed with a dedicated spectral decomposition technique \citep{Coccato2011MNRAS.412L.113C, Katkov2013ApJ...769..105K} that allows one to determine the parameters of the stellar population of both the main and CR stellar disk.
In this paper, we expand the sample of well-studied counter-rotating galaxies by the intriguing galaxy \pgc, a dominant system in a pair outside of any major cluster or group, whose stellar population properties are remarkably different from the other known cases.
Throughout the paper, we have adopted a physical scale of 0.395~kpc~arcsec$^{-1}$ according to NASA/IPAC Extragalactic Database\footnote{\url{http://ned.ipac.caltech.edu/}} for the redshift of $z=0.0195$ and the Planck 2015 cosmology \citep{PlanckCollaboration2016A&A...594A..13P}.

In Section~\ref{sec:observations}, we describe new observations and archival data.
In Section~\ref{sec:analysis}, we analyze our spectroscopic data using different approaches.
In Section~\ref{sec:chemical_evolution} we present our simple model of the history of the metal enrichment.
A discussion and summary are provided in Sections~\ref{sec:Discussion} and \ref{sec:Summary}, respectively.

\begin{figure*}
    \centering
	\includegraphics[width=\textwidth]{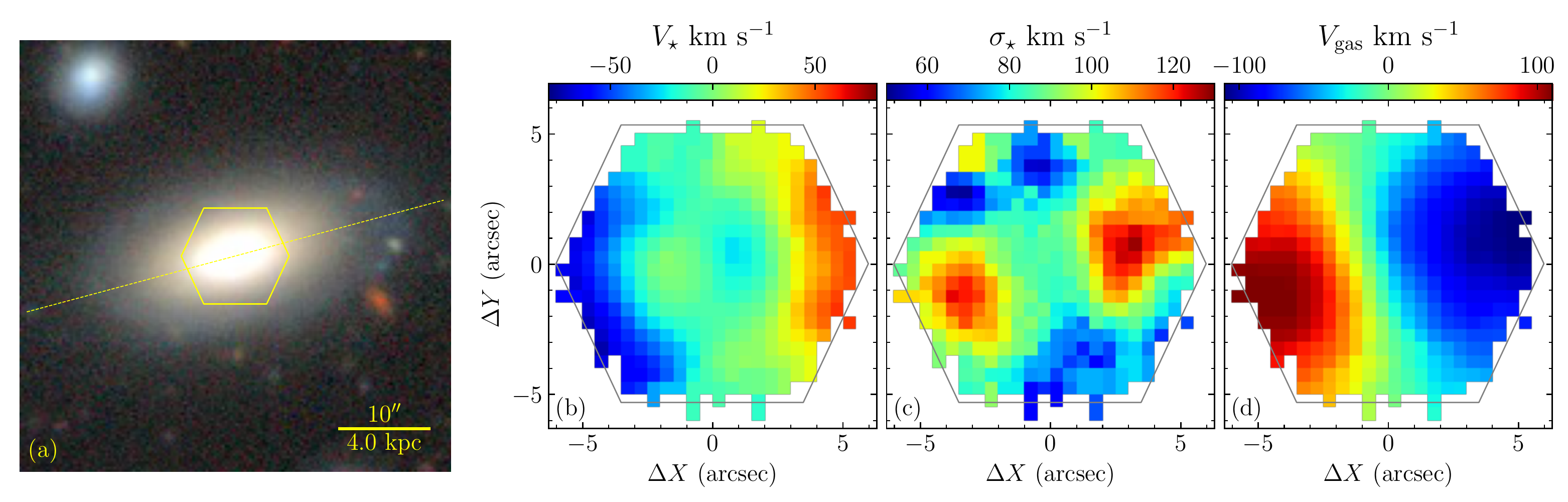}
\caption{A composite RGB image of the PGC~066551 galaxy taken from the DESI Legacy Imaging Surveys (\citet{Dey2019_legacysurveys}, \href{http://legacysurvey.org/viewer?ra=319.7513&dec=-0.9639&layer=dr8&zoom=15}{legacysurvey.org}) is shown in (a).
SALT long slit orientation is shown by dashed yellow line while the hexagon presents MaNGA field of view.
The next three panels show stellar velocity (b), stellar velocity dispersion (c) and ionized gas velocity based on \Ha\ line (d), all taken from MaNGA DR16 Data Analysis Products (DAP).
\label{fig:image}}
\end{figure*}

\section{Observational data}
\label{sec:observations}

\subsection{MaNGA integral field spectroscopy}

The MaNGA survey \citep{Bundy2015ApJ...798....7B} survey contains Integral Field Unit (IFU) \citep{Drory2015AJ....149...77D} observations on the Sloan 2.5 m telescope \citep{Gunn2006AJ....131.2332G} of some 10,000 galaxies in the nearby Universe covering a wide range of stellar mass and colors \citep{Wake2017AJ....154...86W}.
Each galaxy was observed using  2\arcsec\ fibers packed in bundles that varied in diameter from 12\arcsec (19 fibers) to 32\arcsec (127 fibers) covering $(1.5-2.5)\times$ the effective diameter of the target.
{Fiber bundles feed the BOSS spectrographs \citep{Smee2013AJ....146...32S} providing a wavelength coverage of $\lambda\lambda$3,600–10,300~\AA\ with a resolving power of $R\sim2000$, which corresponds to an instrumental dispersion of $\sigma_\mathrm{inst.}\approx75$~\kms\ at 5100\AA\ \citep{Law2016AJ....152...83L}.
To achieve full coverage, each galaxy was observed with three dithered exposures to fill in the gaps between the fibers in a bundle \citep{Law2015AJ....150...19L, Yan2016AJ....152..197Y}.

We inspected data from the DR17 release and identified $\approx$60 galaxies with two distinct off-center peaks in the stellar velocity dispersion maps, indicative of two CR disks (Katkov I. et al. in preparation).
During the inspection process, we intensively used the web-based interactive service for the visualization of MaNGA data (\url{https://manga.voxastro.org}\footnote{\url{https://manga.voxastro.org} is our first prototype service for the visualization of IFU data. \url{https://ifu.voxastro.org} \citep{Katkov2021arXiv211203291K} is the next generation of the service in the early stage of development which includes data from other spectroscopic surveys (SAMI, Califa, Atlas3D).}), which we are developing.
The \pgc\ (MaNGA ID~1-179561) galaxy has one of the best examples of a CR system identified in our sample and therefore was chosen for a separate detailed study.

The MaNGA Data Analysis Product (DAP) map of the stellar velocity (Figure~\ref{fig:image}b) clearly shows that the stellar rotation direction reverses as one moves from the central part to the outer region of the disk.
Two off-center peaks in the velocity dispersion map are another evident feature of the stellar CR disk \citep{Krajnovic2011MNRAS.414.2923K} (Figure~\ref{fig:image}c).
The kinematic position angle, estimated based on the H$\alpha$ velocity map using the \textsc{XookSuut} code \citep{Lopez-Coba2021arXiv211005095L}, yields $PA_\mathrm{kin} = 102.5^\circ \pm 0.2^\circ$.
Meanwhile, the photometrical positional angle derived from the $r$-band SDSS image using the \textsc{isophote} module from \textsc{photutils} package\footnote{\url{https://photutils.readthedocs.io/}} \citep{larry_bradley_2022_6825092} is $PA_\mathrm{phot} = 105^\circ \pm 2^\circ$.
Taking into account the limited MaNGA field-of-view, these estimates appear to be consistent within $\approx1\sigma$.
The similarity of the $PA$ values supports the idea that the gas is settled and rotates in nearly the same plane as the stellar disk.
However, it is not possible to directly compare the rotation of the gaseous disk with that of the stellar one, since the former is a combination of two stellar disks.

\subsection{SALT Long-slit spectroscopy}

To study the properties of CR stars across the entire galaxy disk at a higher spectral resolution and overcome the limitations of the MaNGA field-of-view, we carried out deep long-slit spectroscopic observations of \pgc.
The observations were collected with the Robert Stobie Spectrograph
\citep[RSS;][]{Burgh03,Kobul03} of the Southern African Large Telescope
\citep[SALT,][]{Buck06,Dono06} during the period May 19-31, 2020, in which seven pointings of $2\times1200$s exposures each were made, totaling to 4h40m scientific exposure on target.
We used two configurations of the RSS spectrograph.
We spent 4 hours of total exposure to obtain a deep stellar continuum spectrum using the PG2300 grating with a 1.25\arcsec-wide slit, providing a spectral resolution of $R\approx3200$ in the spectral range of $4500-5500$\AA.
The remaining 40 minutes were used to trace emission lines with a lower resolution of $R\approx2300$ in the $4865-6900$\AA\ range using a PG1300 grating and a 1\arcsec\ wide slit.
The slit was aligned along the major isophotal axis of the galaxy with $PA=285^{\circ}$.
The primary, preliminary and long-slit reduction of the spectra was carried out using the procedure described in \citet{2022AstBu..77..334K}.
We also corrected the science spectra for scattered light in the instrument using the  approach as presented in \citet{Katkov2019MNRAS.483.2413K}.
Reference arc spectra were used to trace the instrumental resolution as a function of wavelength, which is an essential step in the subsequent modeling of spectra.
The reduced spectra for both spectral configurations have a spatial scale of 0.25\arcsec~pixel$^{-1}$, and the spectral sampling for the higher- and lower-resolution setup are 0.32~\AA~pixel$^{-1}$ and 0.64~\AA~pixel$^{-1}$ respectively.
In terms of instrumental velocity dispersion, the spectral resolution is $\sigma_\mathrm{inst}=37$~\kms\ at 5300\AA\ (the location of the prominent \mgb\ absorption feature) for the higher-resolution setup and 48~\kms\ for the lower-resolution setup at 6700\AA\ around the \Ha\ emission line.

\section{Analysis}
\label{sec:analysis}

The main goal of our spectroscopic analysis is to determine the properties of the stellar populations of the CR disks and ionized gas, important for understanding the evolution of that galaxy.
In general, we followed a similar analysis workflow applied earlier to our studies of the CR galaxies IC~719 \citep{Katkov2013ApJ...769..105K} and NGC~448 \citep{Katkov2016MNRAS.461.2068K}.

\subsection{One-component spectral fitting}
\label{sec:spec_nbursts_analysis}

Age and metallicity determine the appearance of a galaxy's stellar spectrum.
The age of the stellar population influences the width and strength of absorption lines, with younger populations exhibiting stronger Balmer lines.
The metallicity primarily affects the number and strength of metal absorption lines, with higher metallicity populations showing more pronounced metal features.
To determine the properties of the stellar population, we applied the full spectral fitting package \nb\ \citep{nburst_a, nburst_b} which models the entire galaxy spectrum and effectively uses the information about the stellar population encoded there.
We used \nb\ with a grid of simple stellar population models (SSP) built with the evolutionary synthesis code \textsc{pegase-hr} \citep{LeBorgne+04} based on the $R=10,000$ empirical stellar library Elodie~3.1 \citep{Prugniel07}.
At each model evaluation during the $\chi^2$-minimization loop, the stellar population spectrum is interpolated from the model grid for a given age $T_\mathrm{SSP}$ and metallicity [Z/H]$_\mathrm{SSP}$ values.
The spectrum is then broadened with stellar line-of-sight velocity distribution (LOSVD), in our case parameterized by a pure Gaussian, and multiplied by a polynomial continuum to account for the difference between the shape of the model spectrum and the observed one due to imperfect spectral sensitivity calibration and effect of dust extinctions.
We used multiplicative Legendre polynomials of the 15th and 20th degrees for the PG2300 and PG1300 setups, respectively.
The model also contains a set of strong emission lines (\Hg, \Hb, [O\iii], [O\oi], [N\ii], \Ha, [S\ii]) possessing the same Gaussian kinematics, but decoupled from the stellar kinematics.
Emission lines are additive components whose weight, i.e. emission line fluxes, are determined as a linear problem solved in every step of a non-linear minimization loop.
The emission line templates as well as a grid of stellar population templates were pre-convolved with wavelength-dependent instrumental resolution before the main minimization loop.
This approach of simultaneously fitting the stellar continuum and emission lines with independent kinematics is similar to the \textsc{gandalf} \citep{Sarzi2017ascl.soft08012S} or recent versions of the \textsc{ppxf} code \citep{Cappellari2017MNRAS.466..798C}.

\begin{figure*}
\centering
\includegraphics[width=0.99\textwidth,trim=0cm 0cm 0cm 0cm,clip]{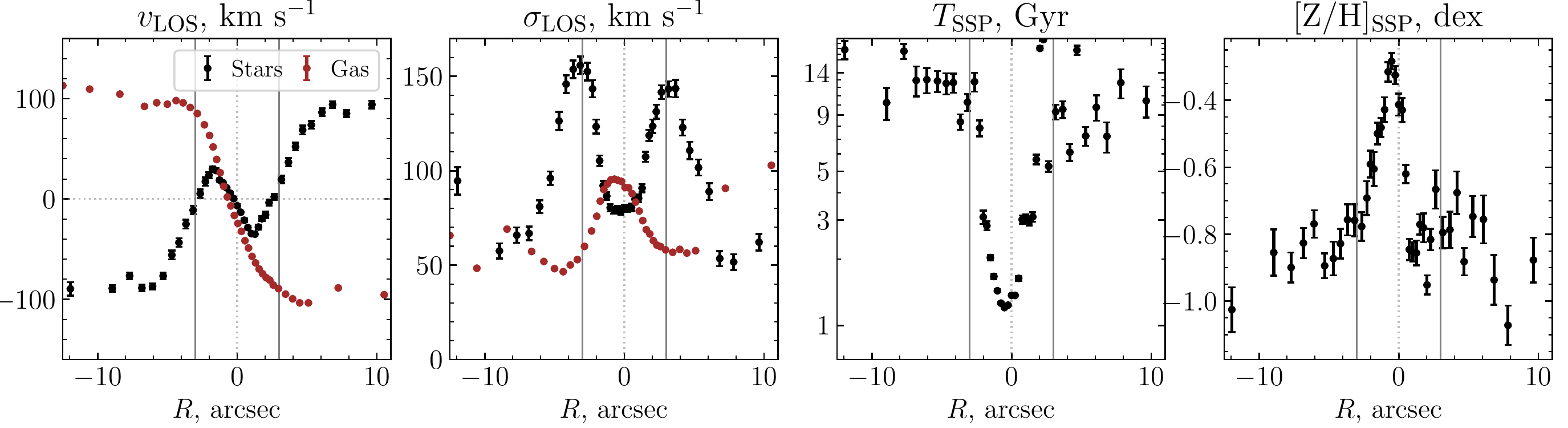}
\caption{Result of one-component analysis of the long-slit spectrum. From left to right panels show line-of-sight velocities, velocity dispersions, SSP-equivalent ages and metallicities. Brown symbols correspond to kinematics of ionized gas.}
\label{fig:one_comp_fitting_profiles}
\end{figure*}

To obtain spatially resolved parameter profiles with reliable measurements, we applied a straightforward adaptive binning scheme to our long-slit data.
This involved increasing the bin size to achieve a minimum signal-to-noise ratio of SNR=10 in the continuum at $5100\pm30$\AA\ in the rest frame.
The radial profiles derived by our single component analysis are shown in Figure~\ref{fig:one_comp_fitting_profiles}.
The kinematic profiles clearly show typical signs of the presence of two CR stellar components \citep{Krajnovic2011MNRAS.414.2923K}, a wavy velocity profile, where in the central region ($R<2\arcsec$) stars rotate in the opposite direction relative to the outer part ($R>5\arcsec$) of the galaxy and a double-peaked velocity dispersion profile.
The two peaks in the velocity dispersion profiles are located at the transition radius $R\approx3-4\arcsec$, where the internal rotation, dominated by one component, switches to the external one, which corresponds to a zero velocity amplitude at that distance.
The same clear dichotomy is observed in the parameters of the stellar population.
The central region is dominated by the younger stellar population, SSP-equivalent age $T_\mathrm{SSP}\approx1.5-2$~Gyr, with metallicity [Z/H]$_\mathrm{SSP}\approx-0.3$~dex.
The outer regions are dominated by a much older stellar population $T_\mathrm{SSP}\approx7-8$~Gyr which is metal-poor [Z/H]$_\mathrm{SSP}\approx-0.8$~dex.

\subsection{Ionized gas}
\label{sec:emission_line_ionized_gas_and_SFR}

The optical spectra of \pgc\ exhibit strong emission lines of ionized gas.
The \nb\ technique used in this study incorporates the contributions of emission lines into the model and provides estimates of line fluxes.
Since the spatial distribution of the emission line fluxes and the stellar continuum are not identical, we applied a different binning schema based on the total \Hb+[O\iii] flux to analyze the emission lines.
This allows us to measure the spatially resolved radial profile of the emission line fluxes and kinematics for the regions extending far beyond the stellar measurements in Section~\ref{sec:spec_nbursts_analysis}.

\begin{center}
    \begin{figure*}
    	\includegraphics[width=1.\textwidth,clip]{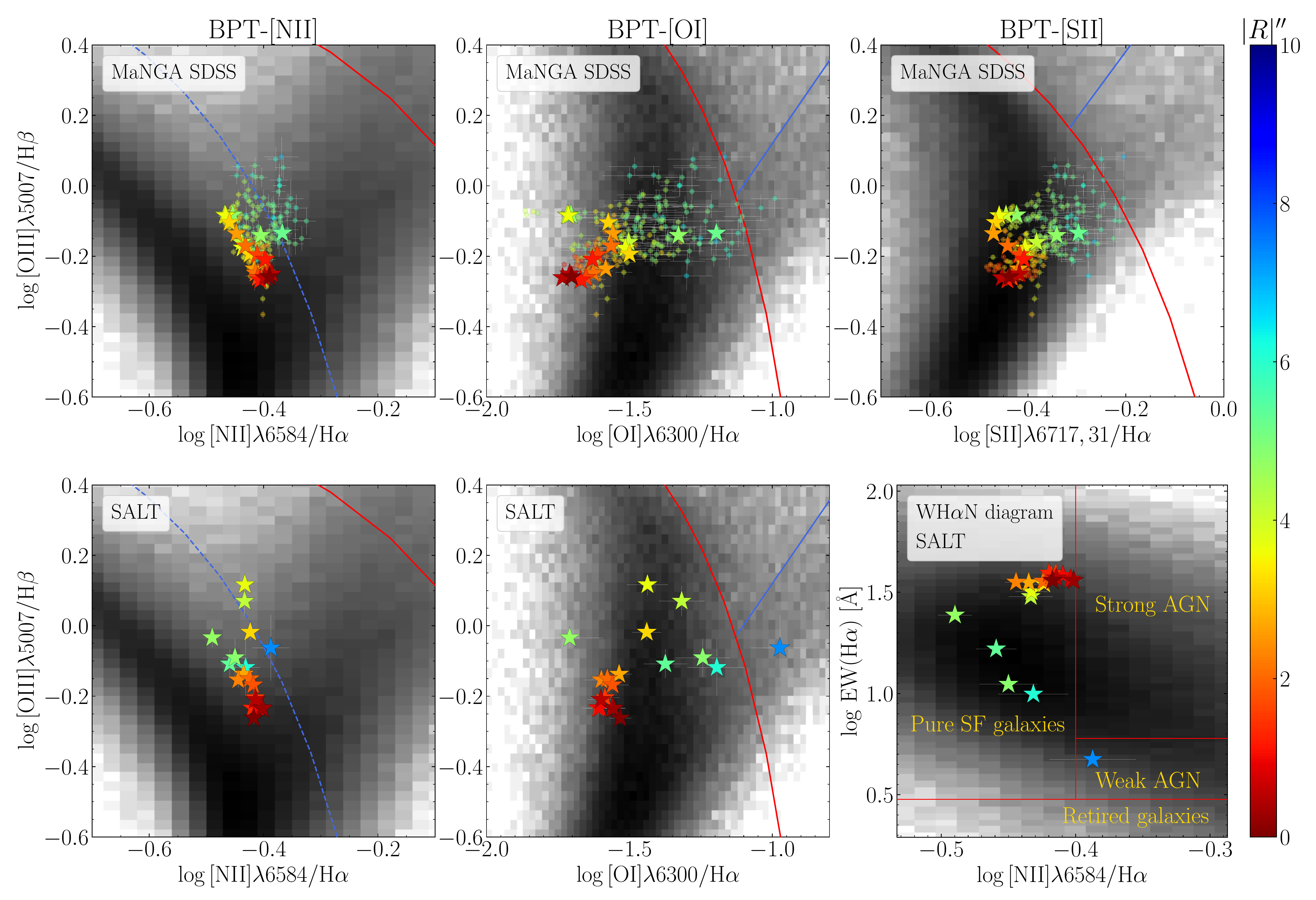}
    \caption{Spatially resolved BPT and WHAN diagrams. The top panel displays emission line measurements derived from MaNGA data, while the bottom panel presents measurements obtained from SALT data. Symbols are color-coded based on their galactocentric distance. In MaNGA plots, spaxels located beyond the major axis are represented as small circles. The blue dashed line in the BPT-NII plot represents the demarcation line introduced by \citet{kauffmann03}, which distinguishes between the star-forming and the so-called composite regions. The additional demarcation lines taken from \citet{kewley01}. They separate composite and AGN regions on the BPT-NII diagram and categorize star-forming, Seyfert/LINER regions in the BPT-OI and BPT-SII diagrams. The bottom right panel also illustrates the WHAN diagram further indicating that all measurements for \pgc\ are situated in the SF region. The distribution of emission line measurements from the RCSED catalog \citep[\url{rcsed.sai.msu.ru},][]{Chilingarian2017ApJS..228...14C} is represented in grey.}
    \label{fig:BPTs}
    \end{figure*}
\end{center}

To identify the gas excitation mechanisms we used both the standard Baldwin-Phillips-Terlevich diagrams \citep[BPT,][]{bpt} with the demarcation lines by \citet{kewley01} and \citet{kauffmann03} and $W_\mathrm{H\alpha}$ versus [N\ii]/H$\alpha$ (WHAN) diagram \citep{CidFernandes2011MNRAS.413.1687C} for both MaNGA and SALT spectra (see Figure~\ref{fig:BPTs}).
Both methods indicate that the dominant excitation mechanism is ionization by young stars with a slight change toward the composite or LINER regions with increasing distance from the galaxy center.
Since no obvious sources of shocks, such as irregular gas motions due to the bar or tidal interaction, are observed, this behavior may be attributed to the increasing relative contribution to ionization by post-Asymptotic Giant Branch (post-AGB) stars of the old stellar population \citep{Yan2012ApJ...747...61Y, Singh2013A&A...558A..43S}.

The dominant excitation mechanism by young stars clearly indicates current star formation.
We corrected emission line fluxes by using Balmer decrement and the \citet{Fitzpatrick1999PASP..111...63F} extinction curve.
Using the total extinction-corrected $H\alpha$ flux enclosed by the MaNGA field-of-view covering $\approx$$2R_\mathrm{eff}$ and the \citet{Kennicutt1998ApJ...498..541K} relation, we estimated the total star formation rate (SFR) as $\approx$0.89~\Msperyr.
This value compares well to $0.93$~\Msperyr\ derived from the broadband spectral energy distribution (SED) fitting and provided by the GALEX-SDSS-WISE Legacy Catalog \citep[GSWLC,][]{Salim2018ApJ...859...11S}. 

We compared these SFR estimations with the expected values for a star-forming galaxy located on the star formation main sequence \citep{Brinchmann2004MNRAS.351.1151B, Salim2007ApJS..173..267S}.
The expected SFR value for a star-forming galaxy with stellar mass of $M_*=2.8\times10^{10}$~\Ms\ (the mass measurement is described in Section~\ref{sec:mass_modelling_RC_decomp}), is 3~\Msperyr\ using the relation from \citet{Salim2007ApJS..173..267S} or $1.4$~\Msperyr\ for time-dependant main-sequence best-fit from \citet{Speagle2014ApJS..214...15S}.
This comparison implies that the current star formation in the \pgc\ is slightly quenched and the galaxy has already moved down from the main star formation sequence (if it was there before).

To discuss the origin of the accreted material that formed the counter-rotating disk, we need to know the gas-phase metallicity.
To determine the ionized gas metallicity, we utilized the O3N2 and N2 calibrations from \citet{2004MNRAS.348L..59P}.
We found that the gas metallicity along the slit and throughout the MaNGA field-of-view shows no variation within the typical uncertainty of 0.1~dex for used calibrations.
The average values of $12+\log \mbox{O/H}$\footnote{Within the range of radii from -8\arcsec\ to 5\arcsec\ where measurements of all emission lines have high signal-to-noise ratio.} for both the O3N2 and N2 calibrations are $8.54\pm0.01$~dex\footnote{Errors here are the RMS of the measurements.} and $8.47\pm0.02$~dex, respectively.
Assuming the solar metallicity to be 8.69~dex \citep{2009ARA&A..47..481A}, our estimates correspond to a slightly subsolar ionized gas-phase metallicity in the range between $-0.2$~dex and $-0.1$~dex.
We also applied the IZI technique \citep{Blanc2015ApJ...798...99B} and found that estimates based on all photoionization model grids except \verb|d13_kappa20| and \verb|d13_kappaINF| \citep{2013ApJS..208...10D} are in agreement with the O3N2 and N2 calibrations.

We also check whether \pgc\ falls on the mass-metallicity relation for star-forming galaxies.
Various metallicity calibrations result in significantly different shapes and positions of the mass-metallicity relation varying by almost 1~dex \citep{Kewley2008ApJ...681.1183K}.
To make a valid comparison, we checked the mass-metallicity relation derived for the same calibrations.
For a galaxy mass of $2.8\times10^{10}$~\Ms\ and the O3N2 and N2 calibrations, we expect a $12+\log \mbox{O/H}$ value of 8.76 and 8.72~dex, respectively.
These values are 0.22-0.25~dex higher than the metallicity in the galaxy and significantly higher than the RMS of the relation itself, which is 0.1~dex \citep{Kewley2008ApJ...681.1183K}.

\subsection{Non-parametric stellar LOSVD}
\label{sec:spec_nonparLOSVD_analysis}

To unveil the two-component kinematics of the stellar population in \pgc, we applied the non-parametric LOSVD recovery approach, which we have successfully used in our past studies \citep[e.g.,][]{Katkov2011BaltA..20..453K,Katkov2013ApJ...769..105K,Katkov2016MNRAS.461.2068K,Kasparova2020MNRAS.493.5464K}.
A spectrum of a galaxy logarithmically binned in wavelength can be modelled as a convolution of a high-resolution spectral template of the stellar population with the stellar LOSVD.
In our approach, we treat the deconvolution problem as an ill-posed linear problem whose solution is the desired LOSVD. We solve the problem using a linear least-squares method with a regularization term that requires smooth solution of the LOSVD and minimization of artifact peaks away from the systematic velocity (so-called ``tail'' regularization).
As a template, we used the unbroadened bestfit model from the \nb\ one-component fitting (Section ~\ref{sec:spec_nbursts_analysis}).
Here we also utilized the latest modification of our method that implements an iterative approach \citep{Gasymov2021arXiv211208386G}. 
Rather than finding the full LOSVD solution, we reformulate the problem to search for corrections required on top of the initial approach of the LOSVD profile.
For the first iteration, we use the Gaussian LOSVD shape from the one-component fitting.
For subsequent iterations, we use the full solution from the previous step.
In this approach, the regularization operates on the correction vector rather than the full LOSVD solution, requiring the corrections to be smoothed.

The resulting stellar LOSVD profiles reconstructed in all spatial bins along the slit are shown in the top panel in Figure~\ref{fig:nonpar_losvd} as a position-velocity diagram and clearly demonstrate the existence of two kinematically distinct components in the stellar population -- the inner one dominates in the central part and co-rotates with the ionized gas, while the main stellar component counter-rotates to the gas and extends to the galaxy outskirts.
We have made the code of our method publicly available as a Python package \textsc{sla}\footnote{\url{https://pypi.org/project/sla/}, \url{https://github.com/gasymovdf/sla}}.
Note that a similar but Bayesian-based approach for non-parametric stellar LOSVD recovery \textsc{bayes-losvd} by \citet{Falcon-Barroso2021A&A...646A..31F} is also available.

\begin{figure}
\includegraphics[width=0.46\textwidth,trim=0.6cm 3.8cm 0.7cm 5cm,clip]{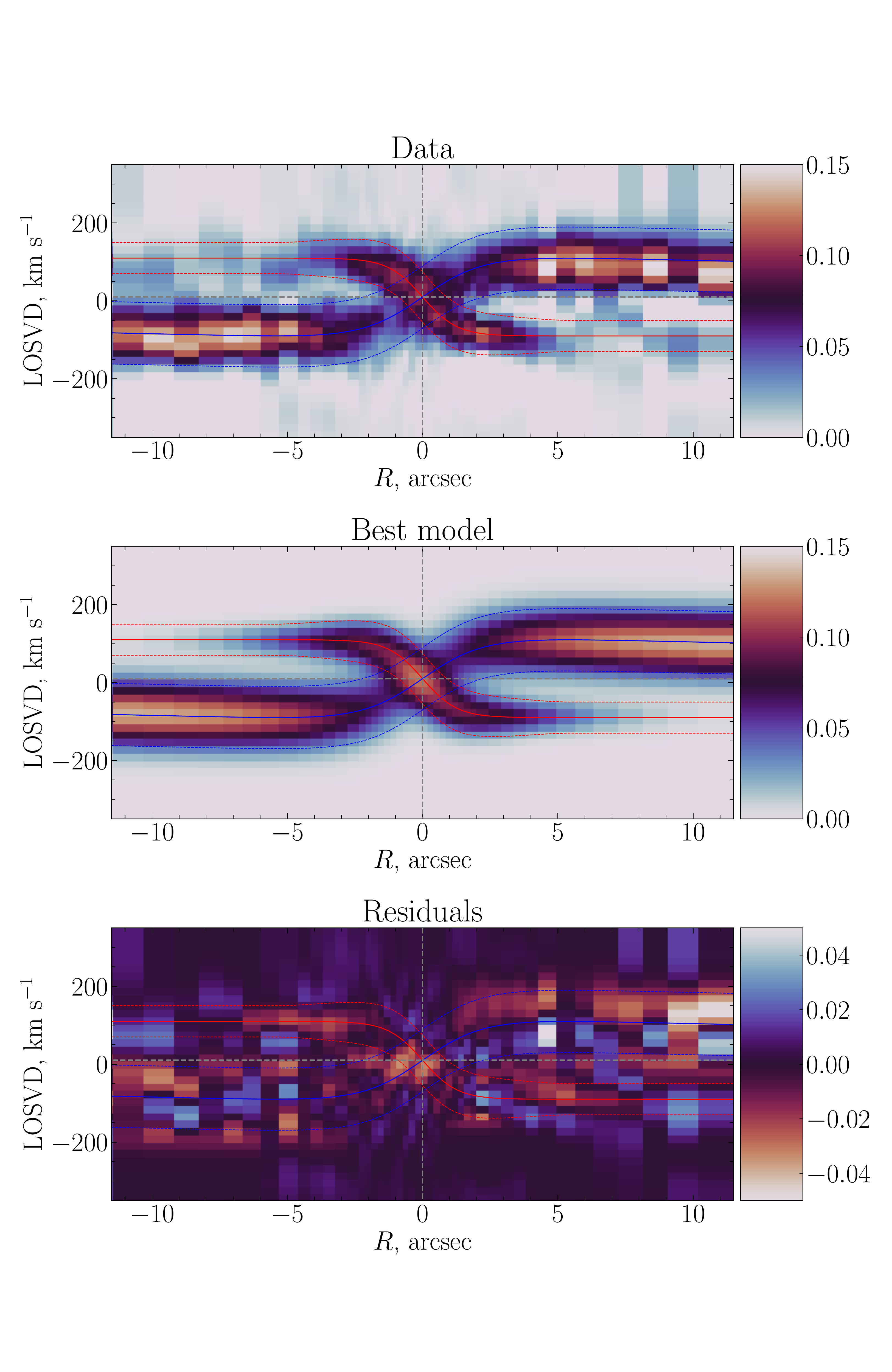}
\caption{Recovered non-parametric stellar LOSVD (top panel) based on the SALT long-slit spectrum. The recovered LOSVD is normalized to unity in each spatial bin. The two counter-rotating component model and its residuals are shown in the middle and bottom panels. Two counter-rotating components are clearly detected, and the CR disk is clearly visible within $\approx5\arcsec$.}
\label{fig:nonpar_losvd}
\end{figure}

To estimate the contribution of the secondary CR stellar component to the integral stellar light, we fit the reconstructed LOSVD. 
Assuming that at any given position of $R$, the LOSVD is parameterized by two Gaussians those parameters sigma $\sigma$, flux $F$ and velocity $v$ are functions of $R$.
For flux and sigma we used exponential decaying laws
$F(R) = F_0 \exp( -|R|/h_f ) $,
$\sigma(R) = \max\{\sigma_0,~\sigma_1 + k |R|\}$,
while for velocity profile we adopted following parameterization:
$v(R)  =  v_\mathrm{max} \left( \tanh(\pi |R|/R_0) + c |R|/R_0\right)$.
The best-fitting LOSVD model and its residuals are shown in the central and bottom panels in Figure~\ref{fig:nonpar_losvd}.
This model suggests that a secondary stellar counter-rotating disk contributes $(42 \pm 2)\%$ to the integral light of \pgc\ galaxy.

One limitation of our LOSVD reconstruction approach is that we use only one stellar template to reconstruct the complex shape of the LOSVD formed by two stellar disks.
It is assumed that both stellar components have comparable stellar population properties and are reliably represented by the same stellar template.
If this assumption is incorrect, this could lead to a distortion of the recovered LOSVD.

\subsection{Two-component spectral decomposition}
\label{sec:spec_2comp_analysis}

To overcome the aforementioned problem of using an identical stellar population template and obtain the stellar population properties for both kinematically distinct components, we performed a two-component parametric fit, also known as a spectral decomposition \citep{Coccato2011MNRAS.412L.113C}.
In this method, the observed spectrum is modeled by two stellar populations, broadened by individual and independent LOSVDs.
We previously implemented this approach based on the full pixel fitting technique \nb\ when studying the stellar counter-rotation phenomena in IC~719 \citep{Katkov2013ApJ...769..105K}, NGC~524 \citep{Katkov2011BaltA..20..453K}, NGC~448 \citep{Katkov2016MNRAS.461.2068K}.
In contrast to those studies, where we masked emission lines, here we incorporated a set of emission line templates into the model.
A decomposition of the spectra from the region $R \approx 2.5\arcsec$ is shown in Figure~\ref{fig:spectral_decomposition}.

\begin{figure*}
    \centering
	\includegraphics[width=0.99\textwidth,trim=0 0 0 0.6cm,clip]{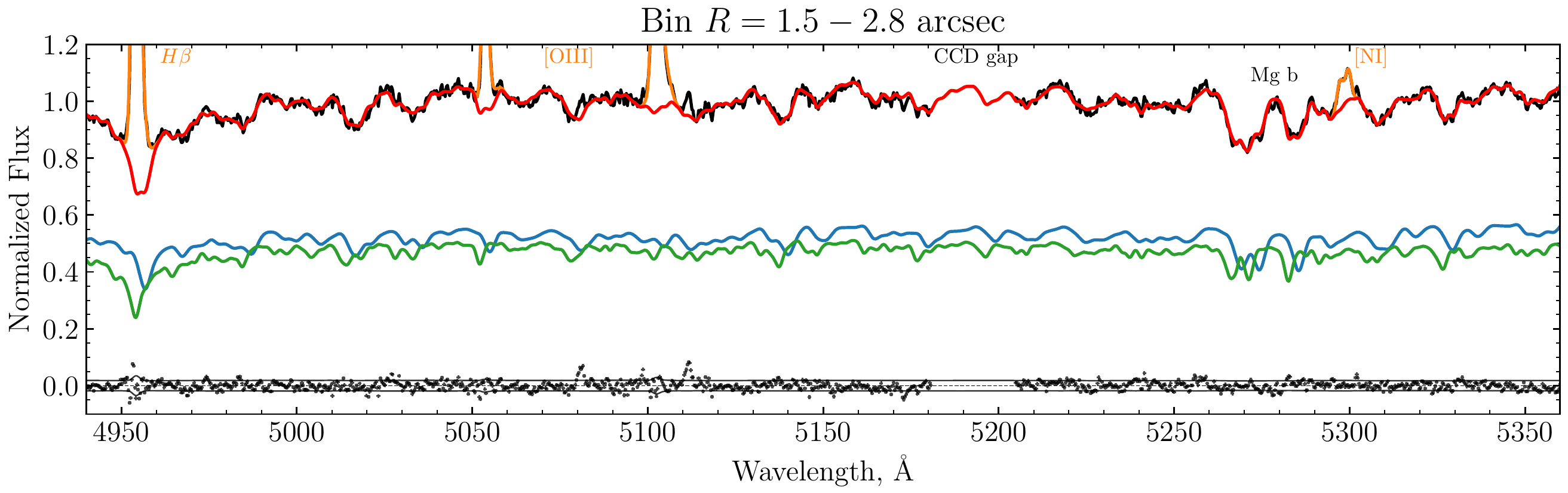}
\caption{Example of the two-component spectral decomposition in the spatial region $R\approx2.5$\arcsec. This figure shows only a fragment of the analyzed SALT spectrum (black line) to demonstrate the fine structure of the absorption lines, in particular \mgb.
Stellar best-fit spectral model shown by red line. Orange lines highlight the emission line model. Main and counter-rotating stellar components are shown in blue and green colors, correspondingly.
Residuals are presented by small dark grey dots bordered by the error level (black lines).}
\label{fig:spectral_decomposition}
\end{figure*}

\begin{figure*}
    \centering
	\includegraphics[width=0.99\textwidth,trim=0.4cm 0.3cm 0cm 1.2cm,clip]{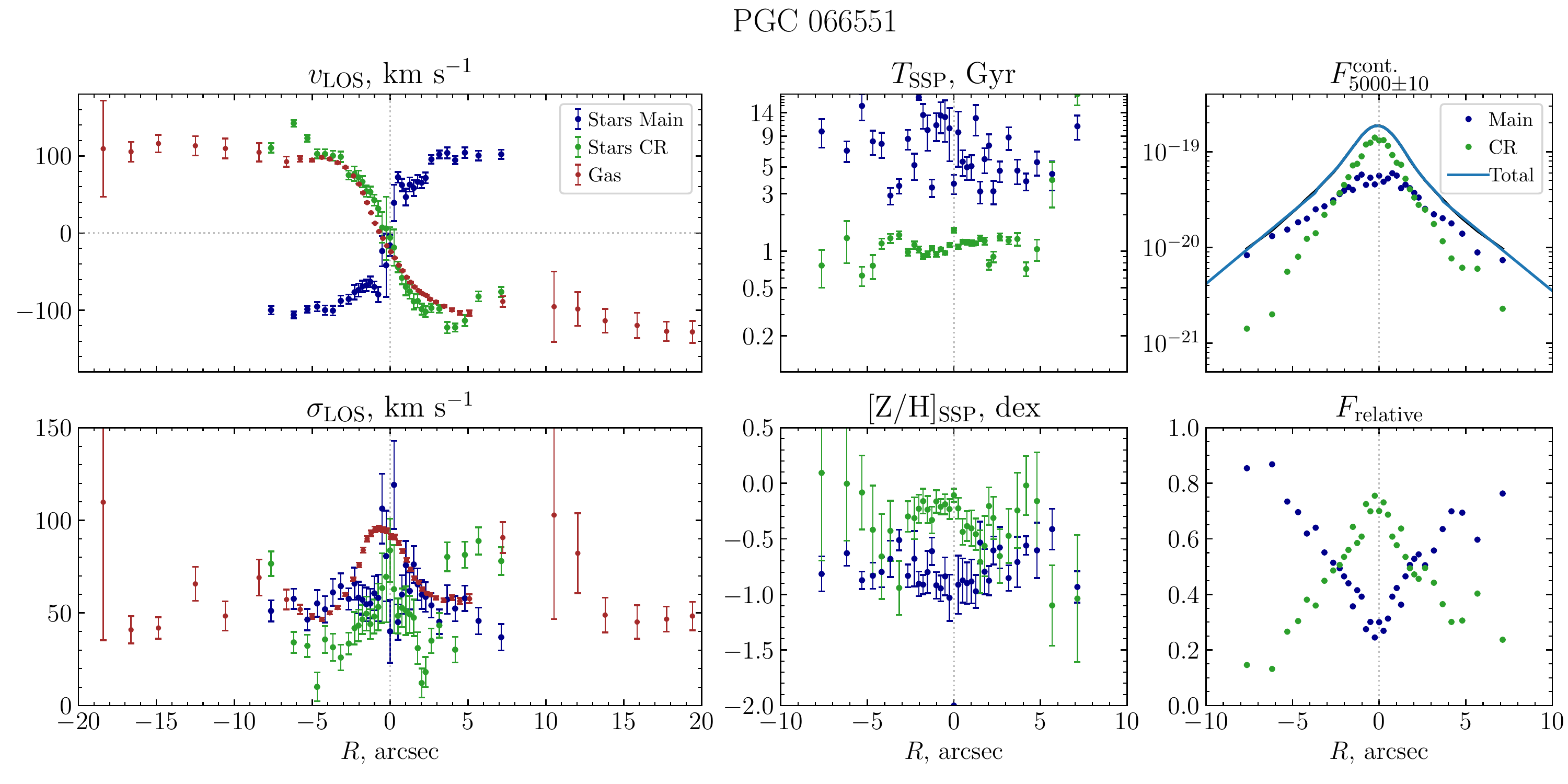}
\caption{Result of two-component SSP analysis. Blue and green points show the main and counter-rotating (CR) components, red points demonstrate gas kinematics. SSP-equivalent ages and metallicities in the middle column clearly suggest the bimodality of stellar populations in \pgc: old main disk and recently formed young CR disks dominated in the central region. The right top panel shows a stellar continuum at $5000 \pm 10 $~\AA\ in the blue line, while the main and CR disk contributions are in symbols. The right bottom plot demonstrates the relative weights of the main and CR component.}
\label{fig:two_comp_profiles}
\end{figure*}

\begin{figure}
    \centering
	\includegraphics[width=0.5\textwidth]{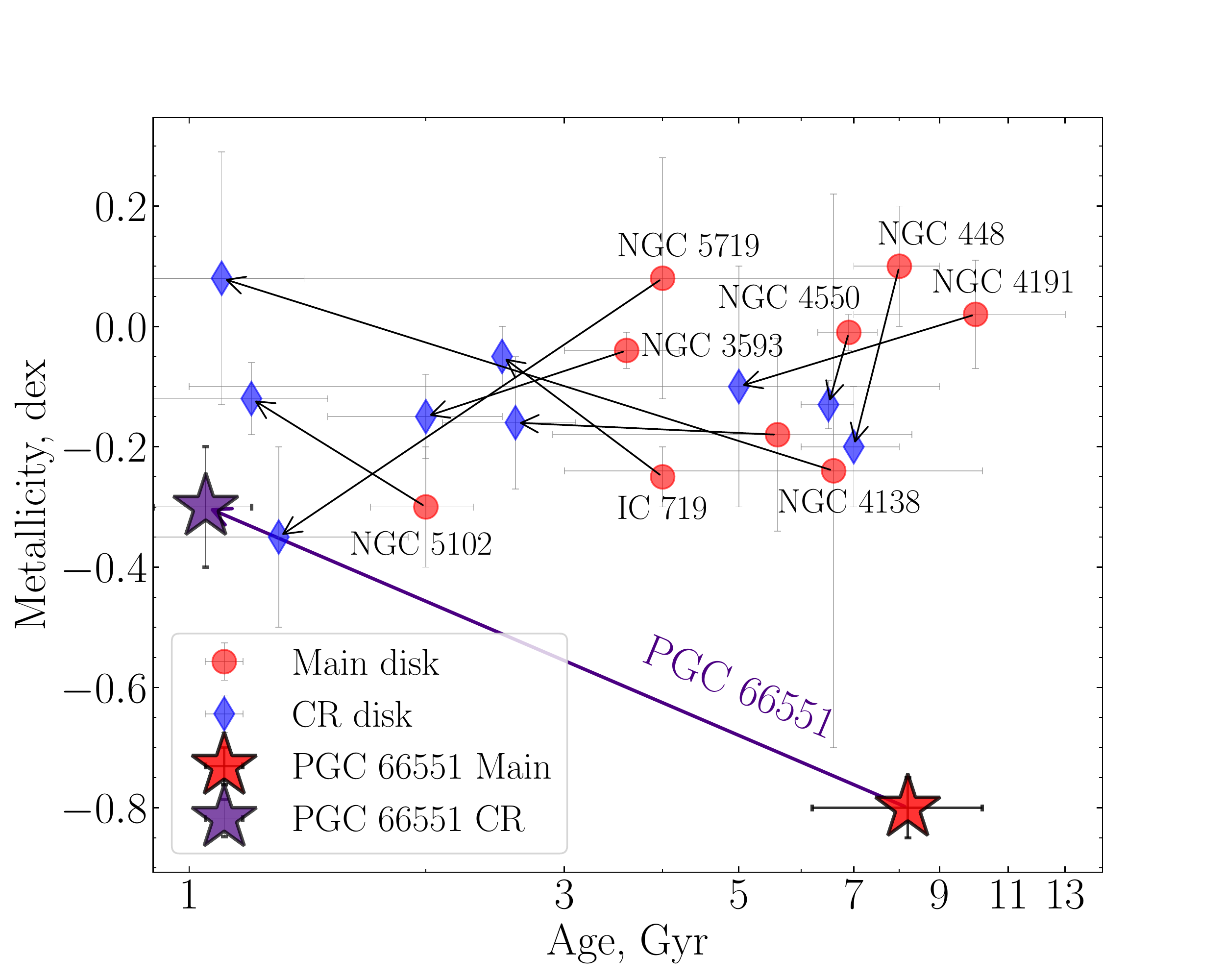}
\caption{Comparison of \pgc\ galaxy with other counter-rotating galaxies where the stellar population properties of the main and CR stellar disks were determined using the spectral decomposition method. The values shown in the figure are SSP-equivalent age and metallicity. Data references: NGC~448 \citet{Katkov2016MNRAS.461.2068K}, NCG~5719 \citet{Coccato2011MNRAS.412L.113C}, NGC~3593 and NGC~4550 \citet{Coccato2013A&A...549A...3C}, IC~719 \citet{Katkov2013ApJ...769..105K}, NGC~4138 \citet{2014A&A...570A..79P}, NGC~4191 \citet{Coccato2015A&A...581A..65C}, NGC~1366 \citet{2017A&A...600A..76M}, NGC~5102 \citet{Mitzkus2017MNRAS.464.4789M}.}
\label{fig:age_met_CRgalaxies}
\end{figure} 

The derived radial profiles of the recovered parameters are shown in Figure~\ref{fig:two_comp_profiles}.
The profiles are less extended compared to the single component analysis because, in this case, we applied a more stringent SNR requirement of 15 during the adaptive binning stage.
They show obvious differences in the line-of-sight velocity, age, and metallicity.
The main and counter-rotating stellar disks, as well as the ionized gas component, exhibit a similar rotation amplitude of $100-120$~\kms\ (not corrected for disk inclination).
However, the main disk counter-rotates with respect to the gas, while the secondary component co-rotates.
The results of the spectral decomposition clearly demonstrate that the counter-rotating stellar disk dominates the integral light in the central 3\arcsec (see right column in Figure~\ref{fig:two_comp_profiles}).
Our assumption that both components have disky morphology is also supported by the radial profile of stellar kinematics.
Both disks have comparable contributions to the integral light at radii $R=\pm3\arcsec$ where two $\sigma$-peaks have been detected in the single component fitting profiles (see Figure~\ref{fig:one_comp_fitting_profiles}).
The main disk is significantly older $T_\textrm{SSP} = 5-10$~Gyr compared to the counter-rotating disk of $\approx 1$~Gyr, indicating that its stars formed much later.
Finally, the metallicity distribution provides a clear distinction between the two disks.
In contrast to most galaxies studied in detail using the spectral decomposition method (see Figure~\ref{fig:age_met_CRgalaxies}), the stellar population of the main stellar disk of \pgc\ exhibits a significantly lower metallicity of [Z/H]$_\textrm{SSP}\approx-0.8$~dex in comparison to the counter-rotating disk which has slightly sub-solar metallicity.
Using the relative weights obtained from the spectral decomposition, we estimated that the contribution of the CR component to the integral light was approximately 50\%, and its contribution to the stellar mass was estimated to be around $50\% (M/L)_\mathrm{CR} / (M/L)_\mathrm{main} \approx 20\%$, based on the mass-to-light ratios derived from the stellar population parameters (see Table~\ref{tab:spec_phot_analysis_pars}).

\subsection{Spectrophotometric fitting}
\label{sec:spectro_photometrical_fitting}

In this Section, we propose a more complex and advanced approach to investigate \pgc\ by simultaneous modeling of both the spectra and broad-band photometry aimed at overcomming the age-metallicity degeneracy and confirming the low metallicity of the main disk, as well as producing a self-consistent stellar population model needed for the analysis of the chemical evolution in Section~\ref{sec:chemical_evolution}.

An increase in the age and metallicity of the stellar population results in the reddening of SED and the effect is particularly strong in the ultraviolet and blue filters \citep{Conroy2013ARA&A..51..393C}.
We have compiled the spatially resolved SED including ultraviolet FUV, NUV bands from Galaxy Evolution Explorer (GALEX) obtained from the MAST archive\footnote{\url{https://mast.stsci.edu/portal/Mashup/Clients/Mast/Portal.html}}, optical $griz$ images from SDSS survey\footnote{\url{https://skyserver.sdss.org}} ($u$-band image was avoided since it is typically faint), $grz$ images from Legacy Survey \citep{Dey2019_legacysurveys}, extracted from cutout service\footnote{\url{https://www.legacysurvey.org/viewer}}, $i$-band CFHT image extracted from CADC\footnote{\url{https://www.cadc-ccda.hia-iha.nrc-cnrc.gc.ca}}, $JHKs$ images from the The VISTA Hemisphere Survey (VHS) retrieved through the ESO Archive Science Portal\footnote{\url{http://archive.eso.org/scienceportal/}}.
We estimated and subtracted sky background level applying $\sigma$-clipping technique\footnote{We used \textsc{sigma\_clipped\_stats()} function from \textsc{astropy.stats} package \url{https://docs.astropy.org/en/stable/stats/}}.
We used azimuthally averaged surface brightness profiles constructed by using \textsc{isophote} module from Astropy-affiliated \textsc{photutils} package.

We used the \textsc{spec+phot} mode of \nb\ \citep{Chilingarian2012IAUS..284...26C, Grishin2021NatAs...5.1308G, Katkov2022A&A...658A.154K_n254} to simultaneously model the optical spectra and broadband SED in two spatial bins $R=\pm2.5\pm0.3$\arcsec\ where the contributions of both disks are comparable.
We initially attempted to use a combination of two SSP models, as in Section~\ref{sec:spec_2comp_analysis}.
However, due to the complex star formation history (SFH) and the current star formation in \pgc\ galaxy, no combination of two SSP models was able to describe the ultraviolet part of the broadband SED.
Then we applied stellar population models with an exponentially decreasing star formation history (hereafter exp-SFH models) defined by the initial epoch of star formation $T_\mathrm{start}$, the exponential decay scale $\tau$, and the mean stellar metallicity.
We used and discussed the same set of stellar population models to analyze stellar population properties of galaxies in the RCSED catalog \citep{Chilingarian2017ApJS..228...14C}.

A spectral grid for exp-SFH models, similar to the SSP grid, was constructed using \textsc{pegase-hr} code \citep{LeBorgne+04} on top of the empirical stellar library Elodie3.1 \citep{Prugniel07} using \citet{Kroupa2001MNRAS.322..231K} IMF.
Photometric models were computed with the \textsc{PEGASE.2} code \citep{Fioc1997A&A...326..950F} using the low-resolution BaSeL synthetic stellar library \citep{Lejeune1997A&AS..125..229L}.
During minimization in the \textsc{NBursts+phot} mode the total $\chi^2$ is equal to a sum of spectral and photometric contributions: $(1-\alpha)^2\chi^2_\mathrm{spec} + \alpha^2 \chi^2_\mathrm{phot}$, where we used equal weights of the contributions $\alpha=0.5$.

While extinction \av\ is a crucial parameter for \textsc{spec+phot} modeling, it is an \textit{external} parameter in the \nb\ package, which means that it is not fitted, but fixed at some predetermined value.
We performed the fitting for a range of \av\ values from 0.0 to 1.5 with a step size of 0.05 and found that the best $\chi^2$ values for both spatial bins corresponded to $A_V = 0.65$~mag.
Thanks to strong emission lines in the analyzed spatial bins we estimated extinction as $A_V = 0.63 \pm 0.24$~mag from the Balmer decrement $H\alpha/H\beta$ which is in remarkable agreement with that derived from the spectrophotometric fitting.

Another constraint on the model parameters we could make from the current SFR estimate, which can be recalculated into $\tau_\mathrm{CR}$ value for any given formation epoch of a CR disk $T_\mathrm{start,CR}$.
The main version of \nb\ can only minimize two-parameter grids of stellar population models.
Hence, we fixed the initial epoch for the main disk as $T_\mathrm{start,main}=13$~Gyr and performed a grid search along all nodes for the epoch of the CR component.
The analysis of both spatial bins revealed that the optimal value for the formation epoch of the CR disk is $T_\mathrm{start,CR}=2$~Gyr.
The completely free parameters remaining in the fitting procedure were the metallicities of both components, the exponential decay scale of the main disk $\tau_\mathrm{main}$, and the kinematics parameters, which we do not discuss here, but are generally consistent with the SSP analysis.
The final parameters of the fitting in two spatial bins are given in Table~\ref{tab:spec_phot_analysis_pars}. Figure~\ref{fig:spec_phot_two_comp_analysis} shows the bestfit model and model components for the bin $R=2.5$\arcsec.

The proposed much more complex and comprehensive analysis preserves our conclusion that the main component of \pgc\ does have a metall poor stellar population.
Compared to the SSP analysis ([Z/H]$_\mathrm{SSP}\approx-0.8$~dex), the metallicity [Z/H]$_\mathrm{expSFH}\approx -0.5$~dex is not as extremely low, but remains way lower than the CR disk metallicity $\approx-0.1$~dex in \pgc\ and strongly lower than that in the main components of the other studied CR galaxies (see Figure~\ref{fig:age_met_CRgalaxies}).

Also in Figure~\ref{fig:spec_phot_two_comp_chi2maps} we illustrate how spectrophotometric fitting breaks the age-metallicity degeneracy.
We draw slice of the $\chi^2$ space on the plane of the main stellar disk parameters [$\tau_\mathrm{main}$, [Z/H]$_\mathrm{expSFH}$] for several fitting variants: i) where only spectral information is analyzed ($\alpha=0.01$), ii) our main model where spectral and photometrical data are used ($\alpha=0.5$) and iii) purely photometrical data fit ($\alpha=1$).
The less elongated and more circular shape of $1\sigma$- and $3\sigma$-contours around minimum in the $\alpha=0.5$ map clearly suggests a lower degree of degeneracy between the age-related parameter $\tau_\mathrm{main}$ and metallicity.

\begin{figure*}
    \centering
	\includegraphics[width=0.99\textwidth]{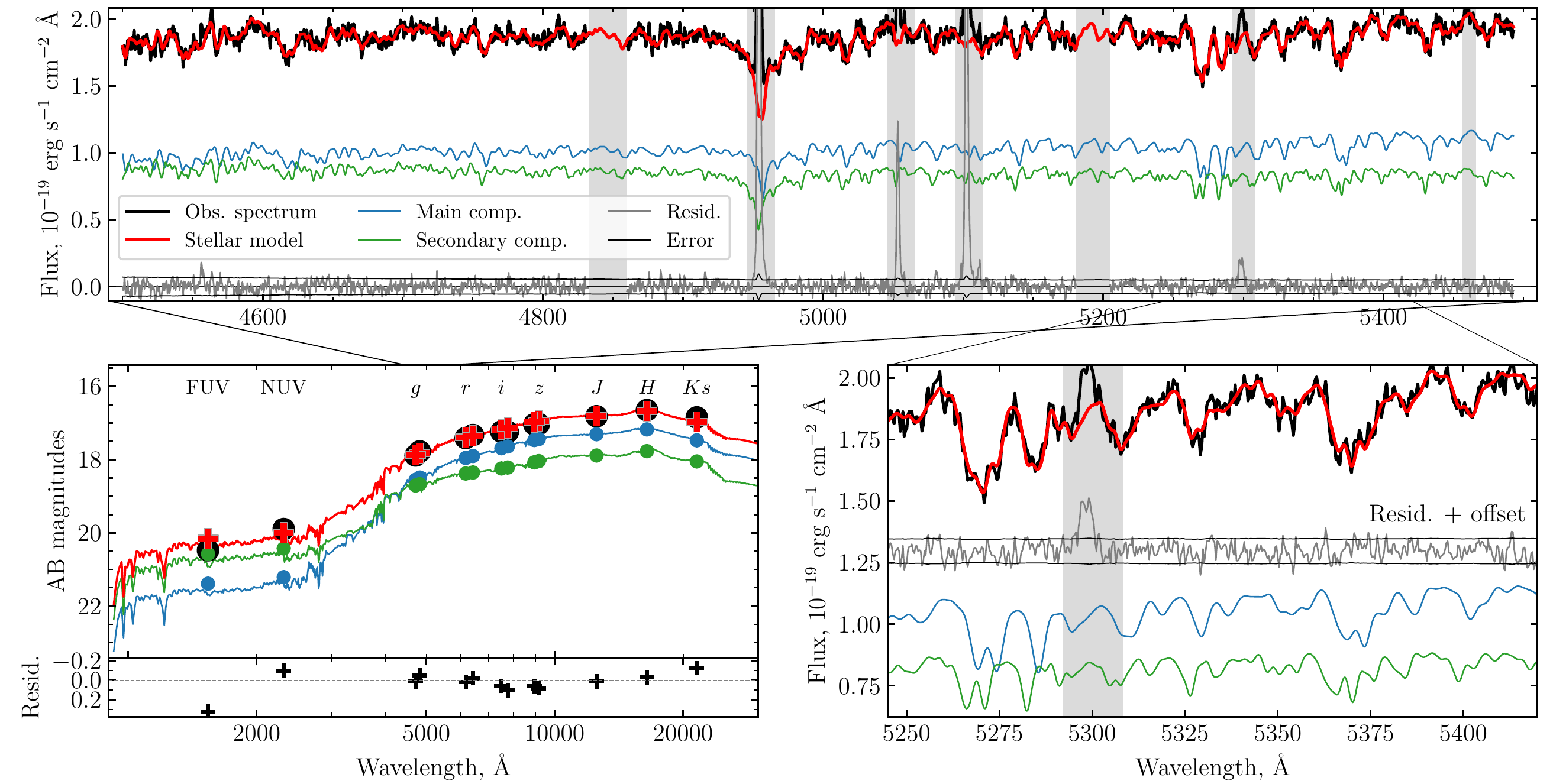}
\caption{Spectrophotometric fitting at spatial bin $R=2.5\pm0.3''$. Top and bottom right panels are identical to that in Figure~\ref{fig:spectral_decomposition}, while the left bottom panel shows broad-band photometrical SED fitting. Black circles are photometrical data for a given spatial bin. Red, blue, and green circles are bestbit model, main and CR components.}
\label{fig:spec_phot_two_comp_analysis}
\end{figure*}

\begin{figure}
    \centering
	\includegraphics[width=\columnwidth]{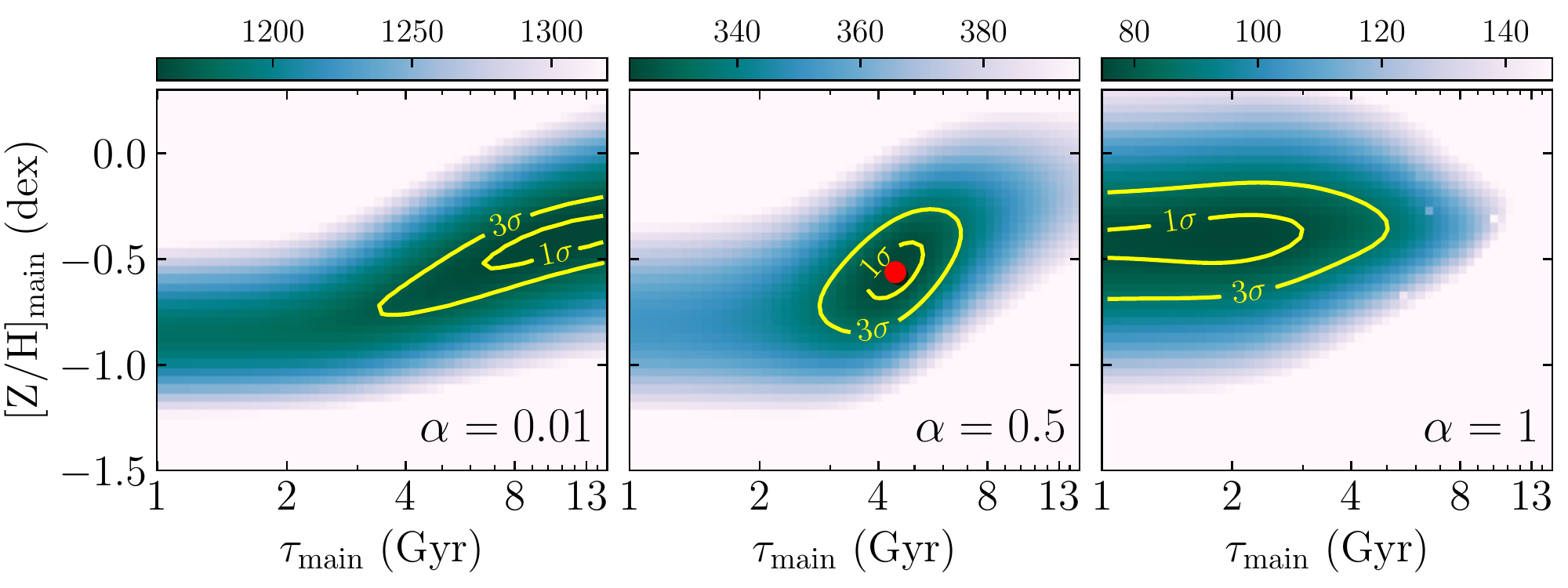}
\caption{
$\chi^2$ scan in the model parameter space for spectrophotometric analysis at spatial bin $R=2.5\pm0.3''$. Maps show $\chi^2$ values in the plane of main stellar disk parameters for a set of different fitting modes: purely spectral fit ($\alpha=0.01$), spectrophotometric fit ($\alpha=0.5$) and purely photometrical fit ($\alpha=1$). Less elongated and more circular contours on the $\alpha=0.5$ map suggest lower degeneracy between parameters. The red circle shows bestfit solution from fitting procedure described in Section~\ref{sec:spectro_photometrical_fitting}.}
\label{fig:spec_phot_two_comp_chi2maps}
\end{figure}

\begin{deluxetable}{ccc}
\tablecaption{Parameters of spectrophotometric analysis\label{tab:spec_phot_analysis_pars}}
\tablewidth{0pt}
\tablehead{
\colhead{Parameter} & \colhead{Main disk} & \colhead{CR disk}
}
% \decimalcolnumbers
\startdata
\hline
\multicolumn{3}{c}{Bin $R=-2.5\pm0.3''$} \\
\hline
$A_V$, mag & \multicolumn{2}{c}{ 0.65 (grid)} \\
$T_\mathrm{start}$, Gyr & 13 (fixed)&  2 (grid) \\
$\tau$, Gyr & $5.25\pm0.53$ &  $0.98$ (fixed) \\
{[Z/H]}, dex & $-0.54\pm0.15$ &  $-0.18\pm0.27$ \\
\hline
\multicolumn{3}{c}{Bin $R=2.5\pm0.3''$} \\
\hline
$A_V$, mag & \multicolumn{2}{c}{ 0.65 (grid)} \\
$T_\mathrm{start}$, Gyr & 13 (fixed)&  2 (grid) \\
$\tau$, Gyr & $4.42\pm0.40$ &  $0.98$ (fixed) \\
{[Z/H]}, dex & $-0.56\pm0.13$ &  $-0.04\pm0.20$ \\
\hline\hline
\multicolumn{3}{c}{Rotation curve decomposition} \\
\hline
$\tau_\textrm{RC}$, Gyr & $4.8$ & $0.98$ \\
{[Z/H]}$_\textrm{RC}$, dex & -0.55 & -0.20 \\
$M/L$, M$_\odot$/L$_\odot$ & $1.69$ &	$0.68$ \\
$\mu_0$ $r$-SDSS, mag & $19.25 \pm 0.07$ & $18.55 \pm 0.16$ \\
$h$, kpc & $1.26 \pm 0.05$ & $0.65 \pm 0.03$ \\
$\Sigma_0$, M$_\odot$/pc$^2$ & $1037 \pm 64$ & $800 \pm 121$\\
$M_\mathrm{disk}$, $10^9$~\Ms & $23.3 \pm 2.0$ & $4.8 \pm 0.8$ \\
\hline
$M_\mathrm{NFW}$, $10^{12}$~\Ms & \multicolumn{2}{c}{$1.6 \pm 0.4$} \\
$M_\mathrm{burk}$, $10^{12}$~\Ms & \multicolumn{2}{c}{$0.8 \pm 0.1$} \\
$M_\mathrm{piso}$, $10^{12}$~\Ms & \multicolumn{2}{c}{$3.0 \pm 0.6$} \\
\enddata
\tablecomments{Rotation curve decomposition section includes adopted parameters of stellar populations ($\tau_\mathrm{RC}$ and [Z/H]$_\mathrm{RC}$) utilized to calculate $M/L$ which is further examined for RC decomposition.}
\end{deluxetable}

\subsection{Rotation curve decomposition}
\label{sec:mass_modelling_RC_decomp}

The SALT long-slit spectrum along the major axis shows that the emission lines trace the galaxy rotation to $20\arcsec\approx8$~kpc, where it has already reached a plateau.
We used this extended rotation curve to estimate the total dynamical mass of the galaxy and the fraction of the dark matter halo.
First, we corrected the determined line-of-sight velocity measurement for the inclination effect by adopting the angle $i=30^\circ$ from the isophote analysis of the $r$-band SDSS image carried out by \textsc{isophote} module from Astropy-affiliated \textsc{photutils} package\footnote{\url{https://photutils.readthedocs.io/}} \citep{larry_bradley_2022_6825092}.
Since the decomposition of the rotation curve is ambiguous \citep[see e.g.][]{Saburova2016MNRAS.463.2523S}, which leads to a strong degeneracy between parameters, we minimized the number of free model parameters to stabilize the solution.
We decomposed the rotation curve into two stellar disks (main and counter-rotating) and a dark matter halo.
To make this analysis consistent with the stellar population analysis, we applied relative weights derived in the spectral decomposition (see Section~\ref{sec:spec_2comp_analysis}) to the $r$-band radial surface brightness profile to calculate individual profiles for both disks.
The profiles were found to be similar to exponential profiles, and therefore, we fit them by exponents and used their parameters to simplify the calculation of their contribution to the rotation curve.
To convert the light profiles into mass profiles, we explored the refined mass-to-luminosity ($M/L$) ratios for adopted stellar population parameters $\tau_\mathrm{RC}$, [Z/H]$_\mathrm{RC}$ from the spectrophotometric analysis (see parameters in Table~\ref{tab:spec_phot_analysis_pars}).
This also yielded in a total mass of $2.8\times10^{10}$~\Ms\ consisting of $M_\mathrm{main}=2.3\times10^{10}$~\Ms\ and $M_\mathrm{CR}=4.8\times10^{9}$~\Ms.
In this way, the contribution from the stellar disks was totally fixed.
We tried to add a bulge component to the model, but the central part of the rotation curve was already well described and there was no empty room for an additional component.
The parameters of the dark matter halo remained free.
Similar to \citet{Saburova2016MNRAS.463.2523S} we considered three types of DM density profiles: Navarro-Frenk-White (NFW) profile \citep{Navarro1996ApJ...462..563N}, the density profile by \citet{Burkert1995ApJ...447L..25B} and the pseudo-isothermal profile (hereafter, piso).
All three profiles are parameterized by central density $\rho_0$ and radial scale of the halo $R_s$.
For convenience, we re-parameterized the profiles in parameters $R_s$ and $V_\mathrm{max}$, where $V_\mathrm{max}$ is a peak halo velocity at radius $R_\mathrm{max}$.
The example of the $\chi^2$-map for the NFW halo in the ($V_\mathrm{max}$, $R_s$) coordinates is shown in Figure~\ref{fig:rot_chi} obviously demonstrates degeneracy even between $V_\mathrm{max}$ and $R_s$.
In the last simplification -- we constrained $V_\mathrm{max}$ to be equal to the value of the rotation curve plateau of 250~\kms\ with only one remaining model parameter $R_s$.
Different halo profiles provided similar mass estimates: $M_\mathrm{NFW} = (1.6 \pm 0.4) \times 10^{12}$~\Ms, $M_\mathrm{burk} = (0.8 \pm 0.1) \times 10^{12}$~\Ms, $M_\mathrm{piso} = (3.0 \pm 0.6) \times 10^{12}$~\Ms.

\begin{figure}
    \centering
	\includegraphics[width=0.46\textwidth]{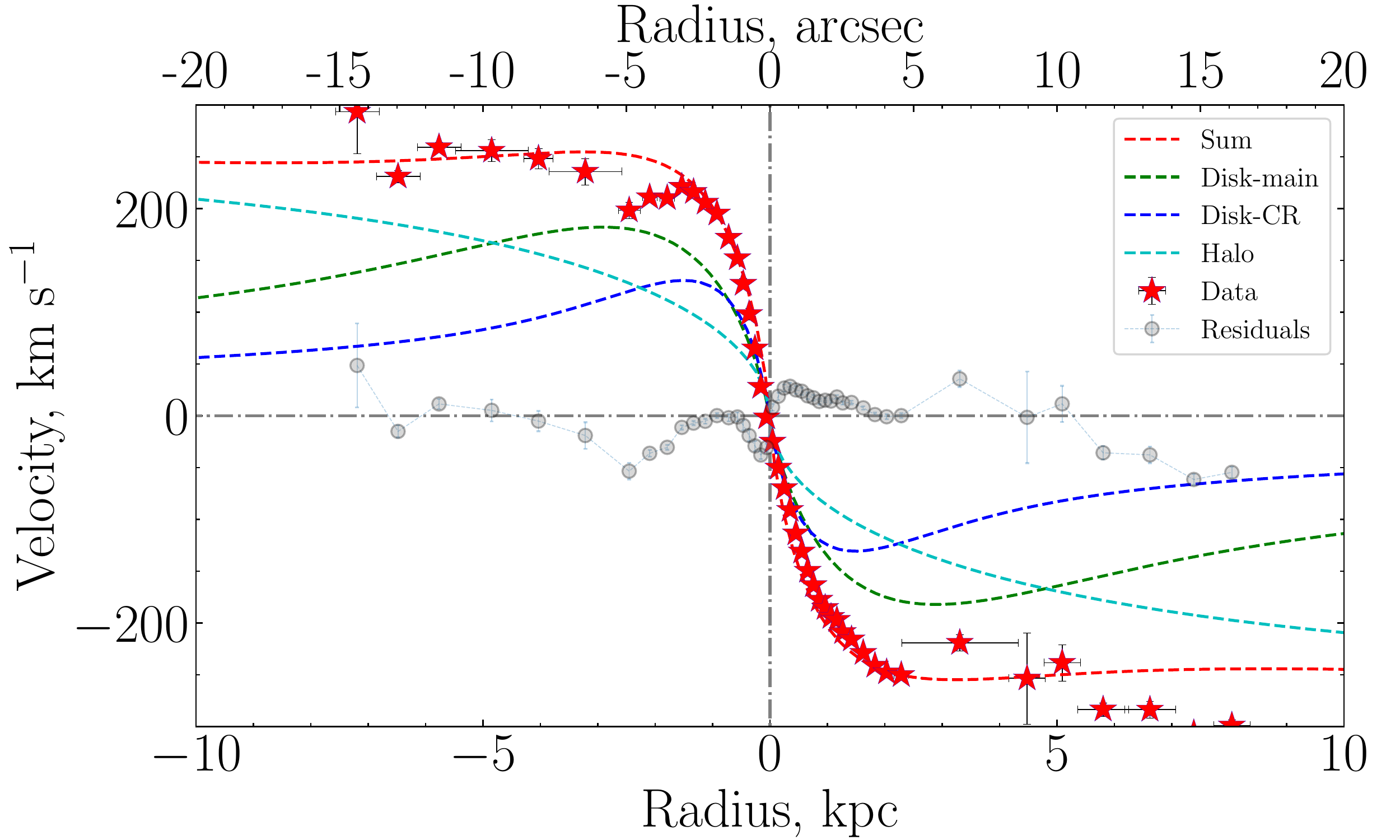}\\
	\includegraphics[width=0.46\textwidth]{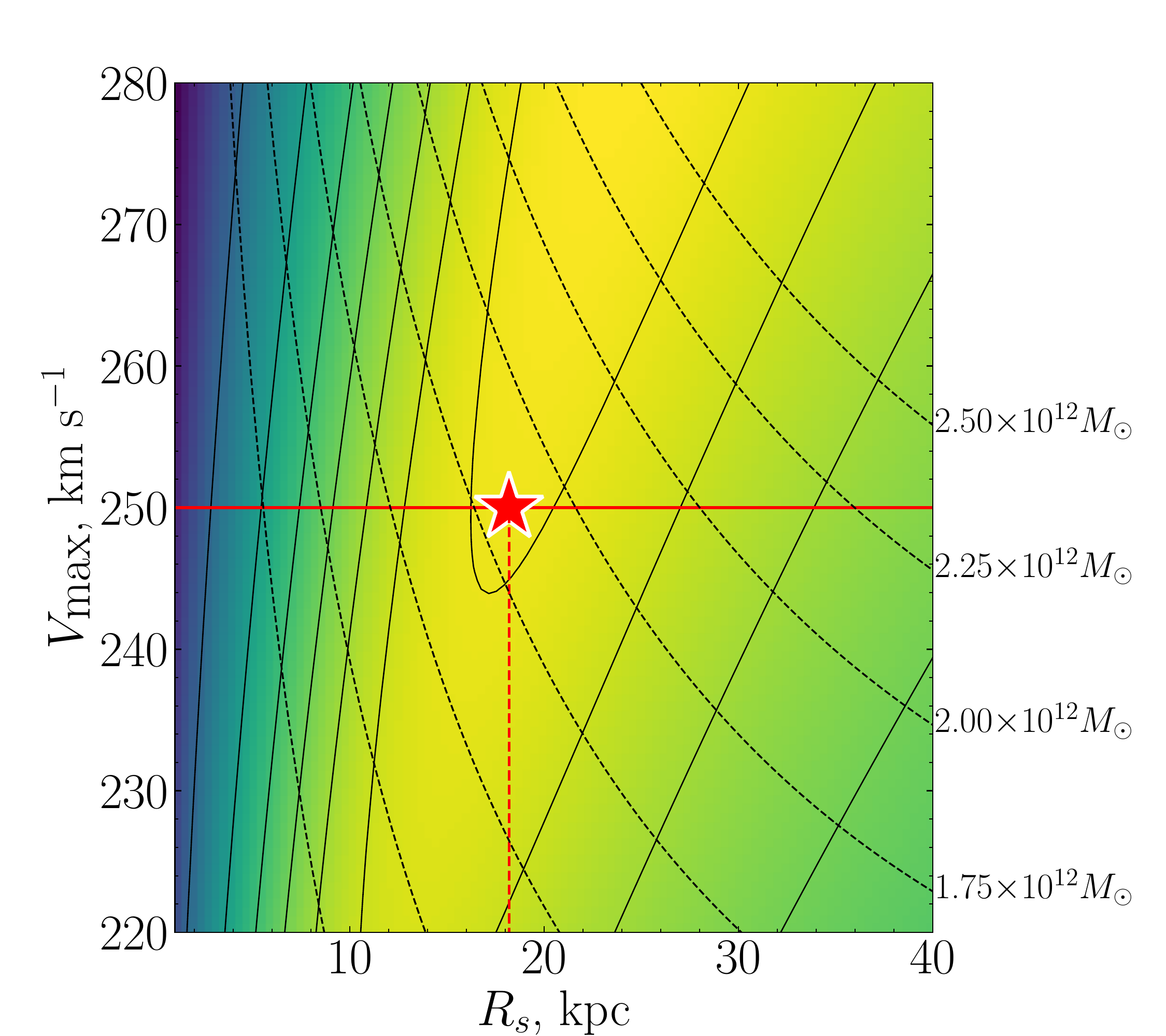}
\caption{Rotation curve decomposition. The top panel displays the rotation curve (magenta stars) corrected for inclination along with the best-fit model (dashed red), disk components (green and blue lines), dark matter halo (cyan line), and residuals (gray dots and line). The bottom panel shows the $\chi^2$-map in the halo parameters coordinates of $V_\textrm{max}$ and $R_s$. The red line indicates $V_\textrm{max} = 250$~\kms, which is adopted from the rotation curve. The red star represents the position of the best-fit value. the black solid lines depict isolines of the $\chi^2$ values, and the black dashed lines demonstrate equal mass of halo.}
\label{fig:rot_chi}
\end{figure}

We found that these halo mass estimates are in perfect agreement with empirical calibration $M_\mathrm{halo} - V_\mathrm{flat}$ \citep{Marasco2021MNRAS.507.4274M} built on the SPARC galaxy sample \citep{Lelli2016AJ....152..157L} and rotation curve decomposition by \citet{Posti2019A&A...626A..56P}. This relation predicts $M_\mathrm{halo}=1.9\times10^{12}$~\Ms\ for velocity plateau of 250~\kms.
Combining $M_\mathrm{NFW}$ and stellar mass (from the GSWLC and our estimates), we obtained the stellar fraction $f_\star=M_\star/f_b M_\mathrm{halo}$, where $f_b$ is a baryon fraction\footnote{$f_b \equiv \Omega_b / \Omega_c = 0.188$ as for adopted cosmology in the \citet{Posti2019A&A...626A..56P} paper.}, in the range of $0.04-0.06$.
In general, this value is in the cloud of the SPARC galaxies on the $f_\star - M_\star$ plot from \citet{Posti2019A&A...626A..56P} (Figure~2) and noticeably lower than the prediction from abundance matching study by \citet{Moster2013MNRAS.428.3121M} (grey stripe on the same Figure~2).
This comparison for \pgc\ implies that, despite its kinematic peculiarities and very unusual way of building its stellar disk, the ratio between the stellar components and dark matter halo is not significantly abnormal.

\section{Model of the chemical evolution}
\label{sec:chemical_evolution}

\subsection{Framework}
\label{sec:framework}

In this section, we construct a framework for modeling the history of the chemical enrichment, which can explain the properties of the stellar population of CR disks and ionized gas in \pgc.
Since we found no significant gradients in the stellar population properties and gas metallicity, we can consider the chemical enrichment as a global process, assuming the entire galaxy as a single well-mixed zone with a uniform star formation history and metallicity.
We exploited \textit{instantaneous recycling approximation} \citep[IRA,][]{Schmidt1963ApJ...137..758S, Pagel1975MNRAS.172...13P, Tinsley1980FCPh....5..287T}, which assumes that stars with $m < 1$~\Ms\ have infinite lifetime, while massive stars $m \geq 1$~\Ms\ instantaneously die as they form enriching the ISM by newly produced metals, which immediately mix into the gas phase and are instantly recycled into new generations of stars.
We constructed the IRA model of chemical evolution following a detailed description by \citet{Spitoni2017A&A...599A...6S} and \citet{Chilingarian2018ApJ...858...63C}.
The model typically includes the following ingredients: SFH $\psi(t)$, initial mass function (IMF), the total mass of the gaseous reservoir, and infall $I(t)$ and outflow $O(t)$ rates.
For the convenience of formulating equations, the following parameters are introduced: $R$ -- the so-called returned mass fraction of gas that returns to the ISM and can be recycled for star formation and $y_Z$ -- the yield per stellar generation that is a fraction of mass in heavy elements synthesized by stars and returned to the ISM through stellar winds and supernova explosions.
Both $R$ and $y_Z$ depend on the IMF and much less on metallicity \citep{Vincenzo2016MNRAS.455.4183V}.
We adopted the following values from \citet{Spitoni2017A&A...599A...6S}
% $R=0.287$ and $y_Z=0.0301$ for \citet{Salpeter1955ApJ...121..161S} IMF and 
$R=0.441$ and $y_Z=0.0631$ for the \citet{Kroupa2001MNRAS.322..231K} and \citet{Chabrier2003PASP..115..763C} IMFs.

To calculate the evolution of metallicity in the ISM $Z = M_Z / M_\mathrm{gas}$ we can define a set of equations.
In the closed-box model, the total mass of metals $M_Z$ in the ISM decreases as some metals are locked up into stars $\dot{M}_Z^-=Z \dot {M}_\mathrm{gas}=-Z \dot{M}_\mathrm{stars}$, while it increases due to enrichment from the evolution of massive stars $\dot{M}_Z^+ = y_Z \dot{M}_\mathrm{star}$.
The balance of mass of heavy elements can be expressed as $\dot{M}_Z = \dot{M}_Z^- + \dot{M}_Z^+ = (-Z + y_z)\dot{M}_\mathrm{star}$.
We defined a set of equations for the evolution of the galaxy mass $M_\mathrm{tot}$, gas mass $M_\mathrm{gas}$ and metallicity, taking into account terms for infall $I(t)$, outflow $O(t)$ and $\dot{M}_\mathrm{star} = (1-R) \psi(t)$, as follows:

\begin{equation}
 \begin{cases}
   \dot{M}_\mathrm{tot}(t) = I(t) - O(t),
   \\
   \dot{M}_\mathrm{gas}(t) = - (1 - R) \psi(t) + I(t) - O(t),
   \\
   \dot{M}_\mathrm{Z}(t) = (-Z(t) + y_Z) (1 - R) \psi(t) + Z_\mathrm{inf} I(t) - Z_\mathrm{out} O(t),
 \end{cases}
 \label{eq:chem_evol_equation_set}
\end{equation}
where $Z_\mathrm{inf}$ and $Z_\mathrm{out}$ are metallicities of the infalling and outflowing gas, respectively. Combining the above equations we have that

\begin{equation}
    \dot{Z} = \frac{(1-R) y_Z \psi(t) + (Z_\mathrm{inf} - Z) I(t) - (Z_\mathrm{out} - Z) O(t)}{M_\mathrm{gas}}
    \label{eq:chem_evol_Zdot}
\end{equation}
We used a simple model where the outflow rate is proportional to the SFR $O(t)=\lambda \psi(t)$ \citep{Matteucci1983A&A...123..121M, Matteucci2012ceg..book.....M}.
The mass-loading factor $\lambda$ represents the efficiency with which the gas is expelled from the galaxy by galactic winds.
Following \citet{Dalcanton2007ApJ...658..941D} we consider the metallicity of the outflowing gas as follows: $Z_\mathrm{out} = \epsilon Z + (1 - \epsilon)Z_\mathrm{SN}$.
The outflow material comes completely from the ISM in the case of the entrainment fraction $\epsilon=1$.
$\epsilon=0$ means that the outflow is dominated by supernova ejecta.
The metallicity of the supernova ejecta is $Z_\mathrm{SN}=\eta y_Z$, where $\eta=4.5-5.2$ for the \citet{Kroupa2001MNRAS.322..231K} IMF \citep{Dalcanton2007ApJ...658..941D}.
For our calculations, we adopted $\eta=4.85$. 
For the infall rate we used the most common exponential formulation $I(t) = A \exp{(-t/\tau_\mathrm{inf})}$ \citep{Matteucci2012ceg..book.....M}.
We applied a fourth-order Runge-Kutta method for Equation~\ref{eq:chem_evol_Zdot} with initial conditions [Z/H](0)=-3 \citep{Bromm2011ARA&A..49..373B}, $M_\mathrm{star}(0)=0$, $M_\mathrm{gas}(0)=M_\mathrm{tot}(0)=M_0$.
Despite the simplicity of such an approach to chemical evolution, it allows one to obtain the physical properties of galaxies at different stages of their lifetime and to model and retrieve from spectra their star formation histories \citep{Grishin2019arXiv190913460G}, as well as additional information on infall and galactic winds 
\citep{Chilingarian2018ApJ...858...63C,Grishin2021NatAs...5.1308G}.

\subsection{Application for \pgc}
\label{sec:chem_model_PGC}

Now that we have defined our chemical enrichment framework, we can adapt it to the context of the studied galaxy.
We are aiming to constrain the properties of the outflowing galactic wind and the metallicity of the infalling gas by reproducing the observables in \pgc. 
We consider two options for constructing the model: 1) first reproducing the properties of the main disk, and then the CR disk, and 2) first reproducing the CR disk, and then the main disk.

Using the exp-SFH scales $\tau_\mathrm{main}$, $\tau_\mathrm{CR}$ from the spectrophotometric analysis (see Section~\ref{sec:spectro_photometrical_fitting}) and stellar mass estimates of the disks we can fully constrain their SFHs $\psi_\mathrm{main}$ and $\psi_\mathrm{CR}$.
We assumed that the main stellar disk was formed from a primary pristine baryonic reservoir of mass $M_0$.
The formation began 13~Gyr ago and proceeded according to exp-SFH with $\tau_\mathrm{main}=4.8$~Gyr and resulted in the disk of stellar mass $M_\mathrm{main}=2.33\times10^{10}$~\Ms.

The fact that the externally accreting material formed a large-scale CR disk means that the host galaxy did not contain a significant amount of gas at the epoch $T_\mathrm{start,CR}=2$~Gyr ago when the accretion started.
Otherwise, the gas clouds would collide, lose angular momentum, and fall toward the galaxy center without forming a CR disk.
% We consider two possible scenarios: (i) natural gas depletion and (ii) an extremely intense episode of AGN activity, which swept out all the gas just before the formation of the CR disk.

Formation of the CR disk started $T_\mathrm{start,CR}=2$~Gyr ago with the accretion of gas with an exponential rate $I(t) \propto \exp{(-t/\tau_\mathrm{inf})}$ and metallicity $Z_\mathrm{inf}$.
From spectrophotometric analysis, an exponential SFH with a scale $\tau_\mathrm{CR}=0.98$~Gyr is required for the current star formation to be at the estimated level of 0.93~\Msperyr.
Initially (at the moment $T_\mathrm{start,CR}$), the CR SFR is highest $\approx$10~\Msperyr\ and no gas, therefore, in order to have a physically meaningful situation, it is necessary to ensure a sufficient gas content.
Therefore, we delayed star formation from starting moment until 50~Myr after the gas infall began.
The change in this time offset does not have a significant effect on the results of modeling.
In our baseline model, we adopted that the exponential scale of infall $\tau_\mathrm{inf}$ is equal to exp-SFH scale $\tau_\mathrm{CR}$.
Later we discuss variations of this choice.
Fixed $\tau_\mathrm{inf}$ allowed us to calculate the normalization of the infall exponential parameterization using the condition that the residual mass of the gas is equal to the mass of atomic hydrogen $M_\mathrm{HI}=1.35\times10^9$~\Ms\ from the H{\sc i}-MaNGA survey \citep{Masters2019MNRAS.488.3396M}.
Here, we assumed that externally accreted stars comprise an insignificant fraction.

The generation of stars formed at any given time inherits the ISM metallicity $Z(t)$.
For any given CEH model of the ISM, a combination of $\psi(t)$ and the mass-to-light ratios from the grid of stellar population models can be used to calculate the luminosity-weighted average metallicity of each disk component.
Requiring these values to be equal to those found from the spectrophotometric fit, [Z/H]$_\mathrm{main}=-0.55$~dex and [Z/H]$_\mathrm{CR}=-0.2$~dex, we can obtain constraints on the parameters of the galactic wind and the metallicity of the infalling gas $Z_\mathrm{inf}$ formed the CR disk.
For a given $\lambda$, to reproduce the low metallicity of the main component, one might want to have a metal-rich galactic wind (low $\epsilon$), but a higher metallicity of the CR disk requires preserving rather than expelling metals (higher $\epsilon$).
Therefore, the properties of both disks set the limits on the entrainment fraction $\epsilon$ of the galactic wind for a given mass-loading factor $\lambda$.

\begin{figure*}
    \centering
	\includegraphics[width=\textwidth, trim=0cm 2cm 2.5cm 2cm]{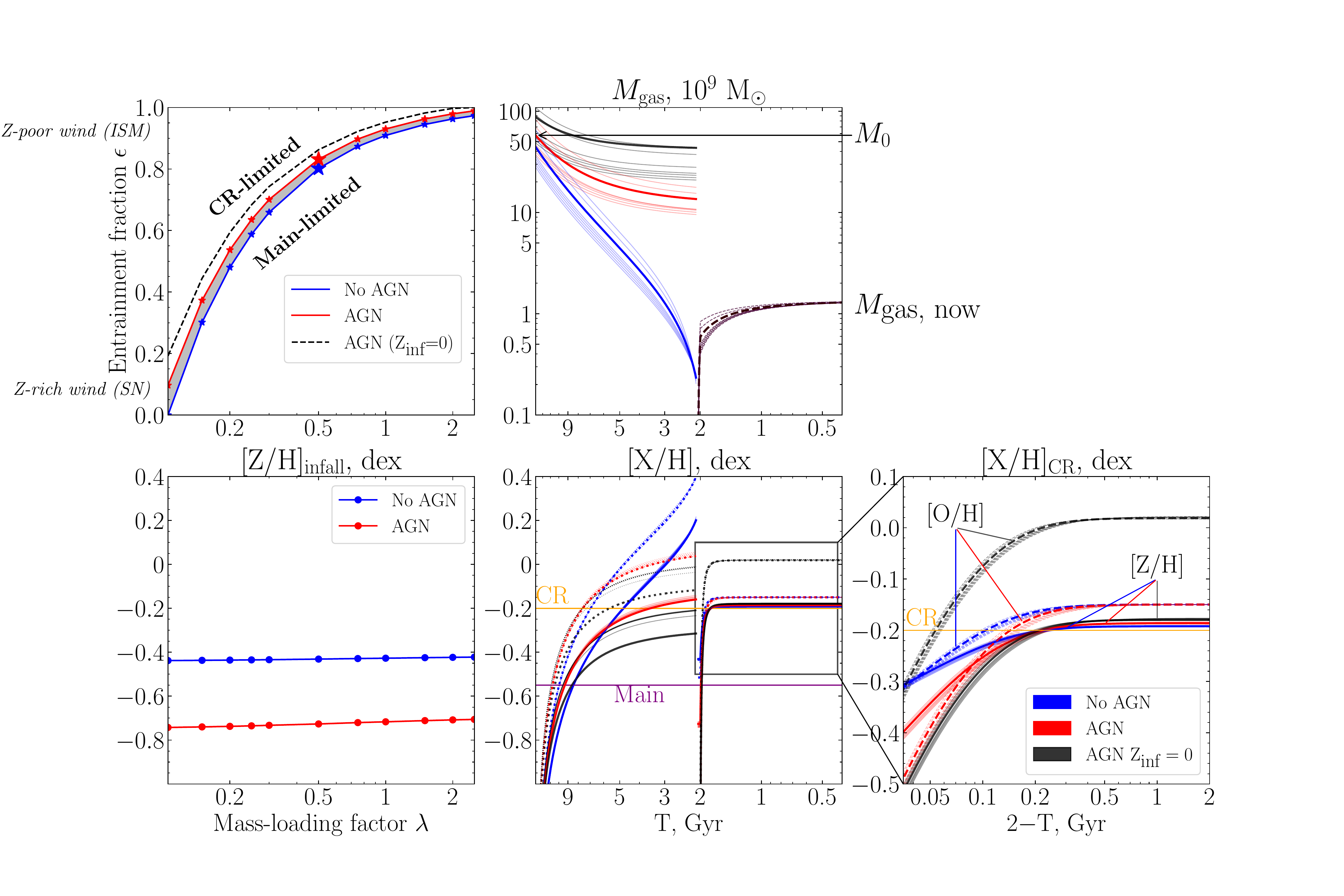}
\caption{Modeling of metal enrichment history. \textit{Top left} panel shows $\epsilon-\lambda$ diagram. $\epsilon$ is entrainment fraction parameter controlling the metal content of the outflowing galactic wind. Red and blue colors correspond to scenarios with and without episode of AGN activity expelled remaining gas at look-back time $T=2$~Gyr. \textit{Bottom left} panel presents metal abundance of the infalling gas formed CR disk. \textit{Top right} panel demonstrates the evolution of total gas mass for different scenarios and mass-loading factors $\lambda$. Tracks for $\lambda=0.5$ are shown by thick lines, thin lines show other $\lambda$ values. Tracks before $T=2$~Gyr corresponds to evolution of the main disk, then the evolution of the CR disk. \textit{Bottom middle and right} plots shows metal enrichment tracks for main and CR disks. Horizontal yellow and purple lines represent metallicities of main and CR disks, which are reproduced in our modeling as luminosity-weighted average values derived from the $Z(t)$ tracks and mass-to-light ratios.}
\label{fig:Chem_evol}
\end{figure*}

We assumed that the outflow process is identical in terms of $\lambda$ and $\epsilon$ during the formation of both disks.
We repeated the below calculations for every $\lambda$ value from the set of 0.1, 0.15, 0.2, 0.25, 0.3, 0.5, 0.75, 1.0, 1.5, 2.0, 2.5.
% First, we focused on the main disk and the scenario without AGN feedback.
First, we focused on the properties of the main disk.
$\epsilon$ is only one free parameter that we can vary to reproduce [Z/H]$_\mathrm{main}=-0.55$~dex.
Low metallicity of the main disk controls the lower boundary in the $\epsilon - \lambda$ diagram shown in blue solid line in Figure~\ref{fig:Chem_evol} (top left panel).
Then, determined $\epsilon$ was used for the history of the enrichment of the CR disk, where we varied the infalling gas metallicity $Z_\mathrm{inf}$ to reproduce [Z/H]$_\mathrm{CR}=-0.2$~dex (solid blue line in the bottom left panel in Figure~\ref{fig:Chem_evol}).

Then we built a model of the metal enrichment focusing on the properties of the CR disk.
In the first step, we considered a variant of the accretion primordial pristine gas assuming $Z_\mathrm{inf}=10^{-3} Z_\odot$ \citep{Bromm2011ARA&A..49..373B}.
Reproducing the stellar metallicity of the CR disk, we obtain $\epsilon$, which is the maximum possible value.
In all other scenarios of pre-enriched gas infall, the $\epsilon$ will be lower, i.e. more metal-rich wind is needed to get the observed stellar metallicity.
This defines the upper boundary shown by the dashed black line in the $\epsilon - \lambda$ diagram.
In this inefficient wind case (higher $\epsilon$), in order to derive the low metallicity of the main disk, we have to increase the initial reservoir of metal-poor gas.
This results in residual gas at the look-back time $T_\mathrm{start,CR}=2$~Gyr which needs to be expelled before the massive accretion of gas forming the CR disk.
Exploring Illustris simulations \citet{Starkenburg2019ApJ...878..143S}, identified two channels responsible for a significant episode of gas removal: an intense burst of AGN feedback and gas stripping occuring during a flyby passage through a massive group environment.
Another possibility, as described by \citet{khoperskov_illustris} and also based on cosmological simulations (IllustrisTNG), suggests gas removal through the capturing and mixing of pre-existing gas with infall.
However, this scenario implies that the mass of infalling gas substantially exceeds the pre-existing one.
Below, we will attribute the episode of gas loss to AGN, implying that this could be any physical process resulting in the effective expelling of residual gas by the time the CR disk formation occurs.}

The oxygen abundance in the ionized gas [O/H]~=~$-0.15$~dex is another observable quantity available to us.
To use it, we added the oxygen enrichment history calculation to the model by solving an identical Equation~\ref{eq:chem_evol_Zdot} assuming the oxygen yield $y_O=0.0407$ \citep{Spitoni2017A&A...599A...6S} instead of $y_Z$.
We found that in the latter scenario, the oxygen abundance appeared to be much higher than the value measured in the galaxy\footnote{Here we assumed zero oxygen enrichment of the infalling gas $O_\mathrm{inf}=0$. Any more oxygen-enriched gas will cause an even larger discrepancy.} (see grey dashed line the bottom right panel in Figure~\ref{fig:Chem_evol}).
To reconcile this discrepancy, the winds must blow metals out more efficiently (lower $\epsilon$).
We minimally reduced $\epsilon$ requiring that the model-predicted oxygen abundance be equal to the observed value, this corresponds to the upper boundary shown by the solid red line in the $\epsilon - \lambda$ diagram.
The metallicity of the CR disk is adjusted according to a variation in $Z_\mathrm{inf}$ (red line in bottom left plot in Figure~\ref{fig:Chem_evol}), while the main disk metallicity is fitted by variations in $M_0$.
In this scenario, the mass of residual gas at $T_\mathrm{start,CR}=2$~Gyr is significantly lower compared to the case of primordial accretion with $Z_\mathrm{inf}=0$ without reproducing oxygen [O/H] abundance in the ionized gas.
Evolution of the total gas mass and metal enrichment history for different $\lambda$ and scenarios are shown in the middle and bottom right panels in Figure~\ref{fig:Chem_evol}.

We chose a very simple exponential infalling rate with $\tau_\mathrm{inf}=\tau_\mathrm{CR}$ as a baseline value.
However, there is no physical reason for this, therefore we tested a set of different infall rates.
We found that the range of possible $\tau_\mathrm{inf}$ values is physically limited.
Indeed, a very high gas infall rate (small $\tau_\mathrm{inf}$) would result in a rapid accumulation of gas in the initial stage and a slow enrichment track, whose averaged value is final metallicity at $T=0$ can be lower than the measured ones.
In the case of prolonged accretion (large $\tau_\mathrm{inf}$), the incoming gas can be totally exhausted and SFR cannot be maintained at the predicted level.
Depending on $\lambda$ both types of problems occur at different $\tau_\mathrm{inf}$.
We chose a secure range of $\tau_\mathrm{inf}$ between $0.6\tau_\mathrm{CR}$ and $1.3\tau_\mathrm{CR}$ for all considered $\lambda$ up to 2.5.
For the lower $\lambda=1$ the range is slightly broader; $\tau_\mathrm{inf}$ can be between $0.5\tau_\mathrm{CR}$ and $1.5\tau_\mathrm{CR}$.
We found that different infall regimes result in different shapes of enrichment tracks.
Namely, the short-scale infall corresponds to gradual enrichment, while the long-scale one to rapid metall enrichment.
In the scenario without expelling gas by AGN at look-back time $T_\mathrm{start, CR}$, the luminosity-weighted metallicity changes insignificantly resulting in very little variation in [Z/H]$_\mathrm{inf}=0.45$~dex.
In the scenario with AGN feedback, [Z/H]$_\mathrm{inf}$ is calculated based on the current ISM metallicity match, which is the last point on the track and is much more sensitive to the shape of the $Z(t)$. 
In this scenario, the [Z/H]$_\mathrm{inf}$ values may vary between $-0.9$~dex for $\tau_\mathrm{inf}=1.5\tau_\mathrm{CR}$ and $-0.5$~dex for $\tau_\mathrm{inf}=0.5\tau_\mathrm{CR}$.
Given the difficulties in measuring absolute metallicity values in the ISM \citep{Kewley2008ApJ...681.1183K}, we tend to consider [Z/H]$_\mathrm{inf}\approx -0.5$~dex as more reliable estimate for infall metallicity, keeping in mind that possibly it can be as low as $-0.9$~dex.

Above, we adopted the atomic hydrogen mass $M_\mathrm{HI}=1.35\times10^9$~\Ms\ as a proxy for the total residual gas mass at $T=0$~Gyr.
However, the unknown contribution of molecular gas can significantly increase this value.
Considering that $M_\mathrm{H2}/M_\mathrm{HI}$ can vary widely \citep{Boselli2002A&A...384...33B, Keres2003ApJ...582..659K} we also re-computed our modeling for an extreme case where the residual mass is equal to $M_\mathrm{gas,now}=10^{10}$~\Ms\ corresponding to $M_\mathrm{H2}/M_\mathrm{HI}=7.4$.
We observed that this modification implies a more massive gas infall and, consequently, a slower enrichment process in the CR disk. Therefore, to achieve the observed metallicity [Z/H]$_\mathrm{CR}$, one needs to start with a higher metallicity of the accreting material, specifically [Z/H]$_\mathrm{inf}=-0.3$~dex in the no-AGN scenario and $-0.7$~dex in the AGN scenario. Additionally, the upper bound of entrainment fractions ($\epsilon$) has increased, making this parameter less constrained.
In addition, we explored a scenario in which residual gas from the main disk formation was expelled (by an AGN event for instance) before the beginning of the CR disk formation ($T_\mathrm{start, CR}$).
For example, the gas expelling at epoch $T=4$~Gyr ago resulted in almost unnoticeable changes in wind parameters, with only slight variations in [Z/H]$_\mathrm{inf}$ decreasing by 0.05~dex.
Both experiments indicate the stability of our simple modeling to various modifications, with one of the crucial output parameters for interpretation, the metallicity of the infall gas [Z/H]$_\mathrm{inf}$, showing only marginal variations.

\section{Discussion}
\label{sec:Discussion}

Comparing \pgc\ with other CR galaxies studied in detail using the spectral decomposition method, the most notable features are the ultra-low metallicity of the main disk and the exceptionally large difference in the ages of the stellar disk populations (see Figure~\ref{fig:age_met_CRgalaxies}).
The first emerged idea was that before the galaxy accreted material to form the CR disk, something stopped the formation of the main disk and kept the disk chemically undeveloped.
Indeed, the exceptionally intensive episode of AGN activity might cause such a dramatic effect \citep{Silk1998A&A...331L...1S, Springel2005MNRAS.361..776S}.
Also, AGN-driven winds inevitably interact with inflows and can drastically reduce accretion rates, affecting star formation in the galaxy \citep{vandeVoort2011MNRAS.415.2782V, Nelson2015MNRAS.448...59N}.
The AGN feedback was explored as one of the mechanisms for gas removal, allowing the subsequent formation of the kinematically misaligned components \citep{Starkenburg2019ApJ...878..143S, Duckworth2020MNRAS.495.4542D}.

The majority of detailed studies of counter-rotating galaxies have addressed the question of formation scenario and source of the externally accreted material that formed the CR disks.
Two main possibilities are usually considered -- accretion from cosmological filaments \citep{Algorry2014MNRAS.437.3596A, khoperskov_illustris} and merger with gas-rich satellites \citep{Thakar1996ApJ...461...55T, Thakar1998ApJ...506...93T, Lu2021MNRAS.503..726L}. 
In our study, we complemented the spectral decomposition approach (and its spectrophotometric extension) by simple chemical evolution modeling which gives some results relevant to this discussion.

In the configuration with AGN feedback episode, our model suggests the metallicity of the infalling gas [Z/H]$_\mathrm{inf}$ to be in the range from $-0.9$~dex to $-0.5$~dex depending on the adopted model of the gas accretion.
The scenario without AGN feedback gives evidence for $-0.45$~dex and is dependent on neither the infall rates nor the wind parameters.
In both scenarios, the expected infall metallicity is higher than that of the near-pristine gas, which we would naively expect in case of accretion from the cosmological filaments.
Direct observational evidence of gas accretion is still extremely challenging and remains  elusive \citep{fox_dave2017}.
However indirect measurements of gas reservoirs in the circumgalactic medium (CGM), probed by absorption features from background quasars, suggest a certain degree of metal enrichment in the CGM, ranging from 2.5\% to 10\% of solar metallicity \citet{Prochaska2013ApJ...776..136P, Lehner2013ApJ...770..138L, Prochaska2014ApJ...796..140P}.
Numerical simulations also suggest that the gas accretion should be metal-poor, but not pristine because of the interaction of cold accretion flows with already enriched galactic halos and outflowing material \citep{vandeVoort2012MNRAS.423.2991V, Hafen2017MNRAS.469.2292H}.
We have a preference towards [Z/H]$_\mathrm{inf}=-0.45$~dex because this value is based on a comparison of the stellar metallicity, which is an average over the entire enrichment history.
[Z/H]$_\mathrm{inf} = [-0.9, -0.5]$~dex seems less reliable, as it depends on the accretion mode and is based on a comparison of the current metallicity in the ISM, which measurements strongly depend on the applied methods \citep{Kewley2008ApJ...681.1183K}.

We also considered gas depletion time $t_\mathrm{dep} = M_\mathrm{gas}/ \mathrm{SFR}$ suggested by different infall scenarios.
More prolonged accretion caused a gradual increase of gas content in the initial stage of CR disk formation which corresponds to a fairly short depletion time $30-130$~Myr depending on the wind parameter $\lambda$.
Typical depletion time in spiral galaxies is $1-2$~Gyr \citep{Bigiel2011ApJ...730L..13B}.
More extreme values can be achieved locally in the star forming regions down to $50-100$~Myr \citep[e.g.,][]{Evans2009ApJS..181..321E, Evans2014ApJ...782..114E, Lada2012ApJ...745..190L}.
More rapid infall leads to faster gas accumulation and as a result, to higher depletion times by factor of two.
Despite still being quite short, the tendency to longer and more realistic depletion time gives support to the rapid rather than prolonged infall.

In symmary, both infall metallicity and regime arguments seem to be more naturally interpreted as a merger with a gas-rich satellite galaxy with the subsequent \emph{in situ} chemical enrichment, rather than a prolonged gas accretion from a cosmic filament.

\section{Summary}
\label{sec:Summary}

In this study, we used deep long-slit spectra obtained with the 10~m class SALT telescope to investigate the galaxy \pgc\ hosting two large-scale counter-rotating stellar disks identified in the SDSS MaNGA survey.
Despite the kinematic peculiarity of \pgc\ which directly points to a very unusual process of building stellar disk, the galaxy generally follows the main scaling relations for disk galaxies.
We found that (i) the metallicity of ionized gas is only slightly lower than expected from the mass--metallicity relation, (ii) the current star formation rate is also slightly suppressed compared to that for star forming galaxies of a given mass residing on the `main sequence of star formation', and (iii) the baryon-to-total mass ratio is comparable to other galaxies but slightly lower than the prediction from simulations.
We employed a novel spectral decomposition technique and its extension for the simultaneous fitting of spectra and broad-band photometry to determine stellar population parameters for both the main and counter-rotating disks.
We demonstrated that simultaneous fitting of spectra and SED can effectively mitigate the age-metallicity degeneracy.

We found a remarkable feature of \pgc\ among the counter-rotating family.
While the young CR disk with sub-solar metallicity contributing $\approx$20\% to the total stellar mass is quite typical for CR galaxies, the main disk populated by old and significantly metal-poor stars has not previously been detected.

We have developed a simple framework for modeling the history of the metal enrichment, which has revealed interesting results for the interpretation of \pgc\ formation.
The metallicities of both disks and current oxygen abundance in the ISM can be reproduced in scenarios with and without AGN feedback.
The assumption of the same properties of the galactic wind during the formation of the main disk and the CR disk makes it possible to constrain wind parameters.
This model also indicates a relatively short time scale of the gas infall, the metal content in it appears to be significantly pre-enriched which is difficult to interpret as evidence for accretion from cosmological filaments.
Therefore, we concluded that the formation of stellar and ionized gas counter-rotation in \pgc\ galaxy is the result of a merger with a gas-rich satellite.

%% Also note that the akcnowlodgment environment does not support long amounts of text. If you have a lot of people and institutions to acknowledge, do not use this command. Instead, create a new \section{Acknowledgments}.
% \begin{acknowledgments}
\section*{Acknowledgments}
We are grateful to the anonymous referee for useful comments and suggestions.
We also thank Anatoly Zasov, Mark Krumholz and Andrea Macci\`{o} for their valuable comments and discussion on this study.
Authors acknowledge the support from the Russian Scientific Foundation grant No. 21-72-00036 and the Interdisciplinary Scientific and Educational School of Moscow University ``Fundamental and Applied Space Research''.
% IC and ER also acknowledge the RScF grant No. 19-12-00281 for supporting the optimization of the \nb\ method for this study.
ER also acknowledges the RScF grant No. 23-12-00146 for supporting the optimization of the \nb\ method for this study.
Spectral observations reported in this paper were obtained with the Southern African Large Telescope (program 2020-1-SCI-002) supported by the National Research Foundation (NRF) of South Africa.
AYK acknowledge the Ministry of Science and Higher Education of the Russian Federation grant 075-15-2022-262 (13.MNPMU.21.0003).
This material is based upon work supported by Tamkeen under the NYU Abu Dhabi Research Institute grant CASS.
Funding for the Sloan Digital Sky Survey IV has been provided by the Alfred P. Sloan Foundation, the U.S. Department of Energy Office of Science, and the Participating Institutions. SDSS-IV acknowledges support and resources from the Center for High-Performance Computing at the University of Utah. The SDSS web site is www.sdss.org.

SDSS-IV is managed by the Astrophysical Research Consortium for the Participating Institutions of the SDSS Collaboration including the Brazilian Participation Group, the Carnegie Institution for Science, Carnegie Mellon University, the Chilean Participation Group, the French Participation Group, Harvard-Smithsonian Center for Astrophysics, Instituto de Astrof\'isica de Canarias, The Johns Hopkins University, Kavli Institute for the Physics and Mathematics of the Universe (IPMU) / University of Tokyo, the Korean Participation Group, Lawrence Berkeley National Laboratory, Leibniz Institut f\"ur Astrophysik Potsdam (AIP), Max-Planck-Institut f\"ur Astronomie (MPIA Heidelberg), Max-Planck-Institut f\"ur Astrophysik (MPA Garching), Max-Planck-Institut f\"ur Extraterrestrische Physik (MPE), National Astronomical Observatories of China, New Mexico State University, New York University, University of Notre Dame, Observat\'ario Nacional / MCTI, The Ohio State University, Pennsylvania State University, Shanghai Astronomical Observatory, United Kingdom Participation Group, Universidad Nacional Aut\'onoma de M\'exico, University of Arizona, University of Colorado Boulder, University of Oxford, University of Portsmouth, University of Utah, University of Virginia, University of Washington, University of Wisconsin, Vanderbilt University, and Yale University.

This work made use of Astropy:\footnote{http://www.astropy.org} a community-developed core Python package and an ecosystem of tools and resources for astronomy \citep{astropy:2013, astropy:2018, astropy:2022}.

This research has made use of the NASA/IPAC Extragalactic Database (NED), which is funded by the National Aeronautics and Space Administration and operated by the California Institute of Technology. This research has made use of NASA's Astrophysics Data System Bibliographic Services.
% \end{acknowledgments}


\vspace{5mm}
\facilities{SALT (RSS), Sloan (MaNGA)}


\software{
    Astropy \citep{astropy:2013, astropy:2018, astropy:2022},
    Photutils \citep{larry_bradley_2022_6825092},
    Pegase.2 \citep{Fioc1997A&A...326..950F},
    lmfit \citep[v1.0.3,][]{Newville2014zndo.....11813N}
}



% \appendix

% \section{Appendix information}


\bibliography{bib}{}
\bibliographystyle{aasjournal}



\end{document}